\documentclass[aps, prd, showkeys, showpacs]{revtex4}

\usepackage[small]{caption}
\usepackage{amsmath,amsfonts,amscd,amssymb,epsf, epsfig}\usepackage{graphicx}

\newcommand{\pa}{\partial}

\begin{document}

\title{Casimir effect of electromagnetic field in Randall-Sundrum spacetime}

\author{L.P. Teo}\email{ LeePeng.Teo@nottingham.edu.my}\address{Department of Applied Mathematics, Faculty of Engineering, University of Nottingham Malaysia Campus, Jalan Broga, 43500, Semenyih, Selangor Darul Ehsan, Malysia. }


\begin{abstract}
We study the finite temperature Casimir effect on a pair of parallel perfectly conducting plates in Randall-Sundrum model without using scalar field analogy. Two different ways of interpreting perfectly conducting conditions are discussed. The conventional way  that uses perfectly conducting condition induced from $5D$    leads to three discrete mode corrections. This is very different from the result obtained from imposing $4D$ perfectly conducting conditions on the $4D$ massless and massive vector fields obtained by decomposing the $5D$ electromagnetic field. The latter only contains two discrete mode corrections, but it has  a continuum mode correction that depends on the thicknesses of the plates. It is shown that under both boundary conditions, the corrections to the Casimir force make the Casimir force more attractive. The correction under $4D$ perfectly conducting condition is always smaller than the correction under the $5D$ induced perfectly conducting condition. These statements are true at any temperature.
\end{abstract}
\maketitle

\section{ Introduction}\label{s1}
Recently there has been considerable interest in studying Casimir effect in Randall-Sundrum brane models \cite{52,53}. A number of works \cite{1,2,3,4,5,6,7,8,9,10,11,12,13,14,15,16,17,36,18,19,20,21,22} have been done from various perspective such as brane stabilization, effective potential, strength of Casimir force, stress energy tensor, etc. The pioneering work of Casimir \cite{23} shows that there is an attractive force of magnitude
\begin{equation}\label{eq7_27_1}F_{\text{Cas}}= -\frac{\pi^2 \hbar c A}{240 a^4}\end{equation} acting between two parallel perfectly conducting   plates with area $A$ and separation distance $a$. In the last thirty years, a number of spacetime models with extra dimensions were proposed to solve different fundamental problems in physics. The prevalence of spacetimes with extra dimensions has motivated research on possible correction to Casimir force acting between parallel plates due to the existence of extra dimensions. Casimir effect on parallel plates in Kaluza-Klein spacetime and Randall-Sundrum spacetime were studied in \cite{24,25,26,27,28,29,30,31,32,33,34,35} and \cite{14,15,16,17,36,18,19,20,22} respectively. Majority of these works considered scalar field instead of electromagnetic field since scalar field is much more easier to deal with. In \cite{14,16,18,24,25},  electromagnetic field were considered but using the scalar field -- electromagnetic field analogy that works in Minskowski spacetime. To yield the correct limit when the size of the extra dimension goes to zero, it was claimed that for the part corresponding to Casimir effect in 4$D$, the Casimir energy due to electromagnetic field is twice the Casimir energy due to   scalar field; whereas for the part corresponding to Kaluza-Klein excitations, the Casimir energy due to electromagnetic field is three times the Casimir energy due to   scalar field. However, it is quite dubious whether such simple relation with scalar field holds especially for warped models. One of the reasons is that the Kaluza-Klein mode masses for scalar field and electromagnetic field are in general different. Another reason is that the perfectly conducting condition for $4D$ massless photons has different generalizations to   higher dimensional massless photons and to $4D$ massive photons.  For simple Kaluza-Klein models such as $M^4\times S^1$, where $M^4$ is the 4$D$ Minkowski spacetime, the scalar field -- electromagnetic field analogy works if one considers $5D$ induced perfectly conducting boundary conditions, but one would obtain three photon polarizations instead of two.  In \cite{26}, a different approach has been proposed for computing Casimir effect on perfectly conducting parallel plates in Kaluza-Klein spacetime $M^4\times S^1$.   In this approach, the Kaluza-Klein zero mode yields the result of Casimir \eqref{eq7_27_1}. For the   Kaluza-Klein excitation modes, it was proposed that they should be treated as Proca fields for massive photons, which have two discrete polarizations and one continuum polarization in the presence of the two perfectly conducting plates.

Recall that  the  spacetime underlying the Randall-Sundrum (RS) model is a 5$D$ anti-de Sitter space (AdS$_5$) with background   metric
\begin{equation}\label{eq7_8_1}
ds^2=g_{\mu\nu}dx^{\mu}dx^{\nu}=e^{-2\kappa|y|}\eta_{ab}dx^{a}dx^{b}-dy^2,
\end{equation} where $\eta_{ab}=\text{diag}(1, -1, -1, -1)$ is the usual 4$D$ metric on the Minkowski spacetime $M^{4}$.  In the following, we will use $\mu,\nu$ for 5$D$ indices taking values from 0 to 4, and $a,b$ for 4$D$ indices taking values from 0 to 3. The extra dimension with coordinate $y$ is  compactified on the orbifold $S^1/\mathbb{Z}_2$. The metric of the underlying Minkowski spacetime depends on the extra dimension through the warped factor $e^{-2\kappa|y|}$, where $\kappa$ determines the degree of curvature of the AdS$_5$ space. There are two 3-branes with equal and opposite tensions, one invisible and one visible, localized at $y=0$ and $y=\pi R$ respectively, where $R$ is the compactification radius of the extra dimension. $\mathbb{Z}_2$-symmetry is realized by $y\leftrightarrow -y$, $\pi R+y\leftrightarrow \pi R-y$.

\begin{figure}[h]\centering \epsfxsize=.3\linewidth
\epsffile{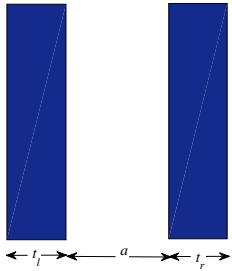}\caption{\label{f1} Two parallel perfectly conducting plates with distance $a$ apart.}\end{figure}
In this article, we    consider the problem of Casimir effect between two parallel perfectly conducting plates in RS  spacetime (see FIG. \ref{f1}).
Two different ways of interpreting perfectly conducting boundary conditions will be discussed. In the first approach, we treat the two perfectly conducting plates as codimension one hyperplanes in 5$D$, and the perfectly conducting boundary condition is the 5$D$ condition directly generalizing the 4$D$ one (see Section \ref{s3} for details).
In the second approach, the electromagnetic field is decomposed into a tower of Kaluza-Klein modes. The Kaluza-Klein zero mode is treated as 4$D$ massless vector field (Maxwell field) and the other Kaluza-Klein modes are treated as 4$D$ massive vector fields (Proca fields) in 4$D$. The perfectly conducting boundary conditions are 4$D$ conditions for massless and massive photons respectively (see Section \ref{s4} for details).  The first approach is in closer spirit to the approach used in \cite{14,16,18,24,25}, and the second approach is the same as the one applied in \cite{26} for the corresponding problem in Kaluza-Klein spacetime.

\begin{figure}[h]\centering \epsfxsize=.6\linewidth
\epsffile{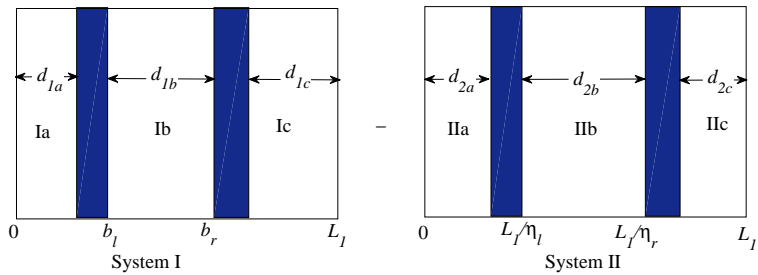}\caption{\label{f2}  The generalized piston approach for renormalization of Casimir energy.}\end{figure}

Recall that the zero temperature Casimir energy and the finite temperature Casimir free energy of a   system at temperature $T$ are defined respectively as
\begin{align*}
E_{\text{Cas}}^{T=0}=&\frac{1}{2}\sum_{\text{modes}}\omega,\\
E_{\text{Cas}}=&-T\sum_{\text{modes}}\ln \left\{\sum_{n=0}^{\infty}\exp\left(-\frac{\omega}{T}\left[n+\frac{1}{2}\right]\right)\right\}=\frac{1}{2}\sum_{\text{modes}}\omega+T\sum_{\text{modes}}\ln \left(1-e^{-\frac{\omega}{T}}\right),
\end{align*}where $\omega$ runs through all nonzero eigenfrequencies of the system. They can be computed using zeta regularization:
\begin{equation}\label{eq7_29_1}\begin{split}
E_{\text{Cas}}^{T=0}=&\frac{1}{2}\left(\text{FP}_{s=-\frac{1}{2}}\zeta(s)+[\log\mu^2]\text{Res}_{s=-\frac{1}{2}}\zeta(s)\right),\\
E_{\text{Cas}}=&-\frac{T}{2} \left( \zeta_T'(0)+[\log\mu^2] \zeta_T(0)\right),
\end{split}\end{equation}
where $\mu$ is a normalization constant, and $\zeta(s)$ and $\zeta_T(s)$ are respectively the zeta functions
\begin{align}\label{eq8_4_2}
\zeta(s)=\sum_{\text{modes}}\omega^{-2s},\hspace{1cm}\zeta_T(s)=\sum_{\text{modes}}\sum_{\ell=-\infty}^{\infty}\left(\omega^2+[2\pi \ell T]^2\right)^{-s}.
\end{align}
To renormalize the Casimir energy, we will use the following setup which is a generalization of the piston approach. Consider a system consists of a large rectangular box $[0, L_1]\times [0, L_2]\times [0, L_3]$ with two plates  placed at $b_l-t_l\leq x^1\leq b_l$ and $b_r\leq x^1\leq  b_r+t_r$, where $t_l$ and $t_r$ are the thicknesses of the plates and $a=b_r-b_l$ is the distance between the plates. Let $E_{\text{Cas}}\left(b_l, b_r, L_1\right)$ be the Casimir energy of this system. Take another reference system where two plates are placed at $L_1/\eta_l-t_l\leq x^1\leq L_1/\eta_l$ and $L_1/\eta_r\leq x^1\leq  L_1/\eta_r+t_r$, where $\eta_l>\eta_r>1$. The renormalized Casimir energy of the parallel plate system is defined as:
\begin{equation}\label{eq7_29_4}
E_{\text{Cas}}^{\parallel}=\lim_{\substack{L_1,b_l,b_r, L_2,L_3\rightarrow\infty\\ \eta_l,\eta_r, a=b_r-b_l\;\text{fixed}}} \left\{E_{\text{Cas}}\left(b_l, b_r, L_1\right)-E_{\text{Cas}}\left(\frac{L_1}{\eta_l}, \frac{L_1}{\eta_r}, L_1\right)\right\}.
\end{equation}Namely, the Casimir energy of the reference system is subtracted before letting $L_1$ goes to infinity keeping the distance between the plates $a$ fixed. See FIG. \ref{f2} for a graphical depiction. This approach is equivalent to the conventional approach of subtracting the Casimir energy in the absence of the plates. For later convenience, let us introduce the notations $d_{1a}=b_l-t_l, d_{1b}=a, d_{1c}=L_1-b_r-t_r, d_{2a}=L_1/\eta_l-t_l, d_{2b}=L_1\left(1/\eta_r-1/\eta_l\right), d_{2c}=L_1-L_1/\eta_r-t_r  $ for the widths of the chambers Ia, Ib, Ic, IIa, IIb and IIc as shown in FIG. \ref{f2}.

The layout of this article is as follows.
In Section \ref{s3}, we consider the Casimir effect on a pair of parallel plates in RS model with $5D$ induced perfectly conducting conditions. In Section \ref{s4}, we consider $4D$ perfectly conducting conditions. Numerical analysis and comparisons of the results are given in Section \ref{s5}. In Section \ref{s6}, we discuss briefly the effect of perturbation by a noncommutativity parameter to the sign of Casimir force.

In this article, we use units where $\hbar=c=k_B=1$.
\section{Casimir effect with 5D induced perfectly conducting conditions}\label{s3}
In this section, we treat the two parallel plates as codimension one hyperplanes in  the 5$D$ spacetime, and impose the 5$D$ induced perfectly conducting boundary condition on the   plates.
In 5$D$ vacuum, the bulk action of the electromagnetic field $F_{\mu\nu}=\pa_{\mu}V_{\nu}-\pa_{\nu}V_{\mu}$ is given by
\begin{equation}\label{eq8_2_2}
S=-\frac{1}{4}\int d^4x  \int dy \sqrt{|g|}  F_{\mu\nu}F^{\mu\nu},
\end{equation}where $g=\det g_{\mu\nu}=e^{-8\kappa|y|}$.
As usual, there is a gauge degree of freedom given by
\begin{equation}\label{eq7_28_1}V_{\mu}\mapsto V_{\mu}+d\varphi\end{equation} for an arbitrary function $\varphi$. The equation of motion is
\begin{equation} \label{eq7_28_4}
  \sqrt{|g|}^{\;-1} \pa_{\mu}\left( \sqrt{|g|}F^{\mu\nu}\right) = \sqrt{|g|}^{\;-1} \pa_{\mu} \sqrt{|g|}g^{\mu\kappa}g^{\nu\eta}\left( \pa_{\kappa}V_{\eta}- \pa_{\eta} V_{\kappa} \right) =0.
\end{equation}The $5D$ perfectly conducting boundary condition is given by \cite{37}:\begin{equation}\label{eq7_28_2}\left.n^{\mu}F^*_{\mu \nu_1 \nu_{2}}\right|_{\text{interface}}=0.\end{equation}
Here $n^{\mu}$ is the spacelike vector normal to the plates, and
$\displaystyle F^*_{\mu_1 \mu_2 \mu_{3}}=\varepsilon_{\mu_1\mu_2\mu_{3}\nu\lambda}F^{\nu\lambda}$ is the $3$-form dual to $F$. Choosing $x^1$ as the direction normal to the plates, this condition reads as
\begin{equation}\label{eq8_2_1}
\left.F_{\mu\nu}\right|_{\text{interface}}=\left.\pa_{\mu}V_{\nu}-\pa_{\nu}V_{\mu}\right|_{\text{interface}}=0\hspace{1cm}\text{for}\quad\mu,\nu\neq 1.\end{equation}
This is the generalization of the $4D$ perfectly conducting condition for massless vector field where the transverse components of the electric field and the normal component of the magnetic field vanishes on the interfaces between the vacuum and the plates.

For the one-form $V_{\mu}dx^{\mu}$, there are two different orbifold boundary conditions with respect to the $\mathbb{Z}_2$ symmetry \cite{39}. One is $V_{\mu}dx^{\mu}$ is even which implies that $V_{a}$ is even and $V_y$ is odd, and the other one is $V_{\mu}dx^{\mu}$ is odd which implies that $V_{a}$ is odd and $V_y$ is even. In the second case, there is no Kaluza-Klein zero mode \cite{38}.  Therefore we will only consider the first case where $V_{a}$ is even and $V_y$ is odd.

To fix the gauge, we impose the Lorentz condition:
\begin{equation}\label{eq7_28_6}
0=\sqrt{|g|}^{\;-1}\pa_{\mu}\sqrt{|g|}V^{\mu}=e^{2\kappa|y|}\left(\pa_0V_0-\pa_1V_1-\pa_2V_2-\pa_3V_3-e^{2\kappa|y|}\pa_ye^{-4\kappa|y|}V_y\right).
\end{equation}As in the 4$D$ case, this does not fix the gauge uniquely. One still has the freedom of adding to $V_{\mu}dx^{\mu}$ an exact one-form of the form $\pa_{\mu}\phi dx^{\mu}$ which satisfies the equation $$ \left(\pa_0^2-\pa_1^2-\pa_2^2-\pa_3^2-e^{2\kappa|y|}\pa_ye^{-4\kappa|y|}\pa_y\right)\phi=0.$$
Under the Lorentz condition \eqref{eq7_28_6}, the equation of motion for $V_y$ (eq. \eqref{eq7_28_4} with $\nu=y$) is
$$\left(\pa_0^2-\pa_1^2-\pa_2^2-\pa_3^2- \pa_ye^{2\kappa|y|}\pa_ye^{-4\kappa|y|} \right)V_y=0.$$Therefore it is consistent to choose  $\phi$ so that $\pa_y \phi=-V_y$. In other words, we can impose the axial gauge $V_y=0$. The Lorentz condition \eqref{eq7_28_6} then becomes
\begin{equation}\label{eq7_28_7}
\pa_aV^a= e^{2\kappa|y|}\left(\pa_0V_0-\pa_1V_1-\pa_2V_2-\pa_3V_3\right)=0.
\end{equation}
 Such $\phi$ is determined up to a function $\psi$ independent of $y$ and satisfying $\eta^{ab}\pa_a\pa_b\psi=0$. It should be remarked that axial gauge can be imposed here because we only consider fields that are even under $\mathbb{Z}_2$ symmetry. For the Kaluza-Klein spacetime  $M^4\times S^1$, one cannot impose the axial gauge \cite{38}.

  Under the Lorentz gauge \eqref{eq7_28_7} and the axial gauge $V_y=0$, the equation of motion \eqref{eq7_28_4} can be simplified to \cite{39}:
\begin{equation}\label{eq7_28_8}
\begin{split}
\left(\eta^{bc}\pa_b\pa_c -\pa_y e^{-2\kappa|y|}\pa_y\right)V_a=0.
\end{split}
\end{equation}Consider solutions of \eqref{eq7_28_7} and \eqref{eq7_28_8} which have the form
$$V_a(x,y)=A_a (x)f(y),$$where $f(y)$ is an even function under the $\mathbb{Z}_2$ symmetry. Here $x=(t,x^1,x^2,x^3)$.

\vspace{0.2cm}
\noindent
\textbf{A. Kaluza-Klein zero mode.}\; The Kaluza-Klein zero mode is the mode corresponding to $f(y)=f_0(y)=$ constant. In other words, $V_{\mu}dx^{\mu}$ is independent of $y$. In this case, there is still a gauge degree of freedom which can be fixed by the condition $V_0^0=0$.

\vspace{0.2cm}
\noindent
\textbf{B. Kaluza-Klein nonzero modes.}\; For the nonzero modes, $f(y)$ is not a constant. \eqref{eq7_28_8} implies that $A_a$ and $f$ have to satisfy the system:
\begin{equation*}
\begin{split}
\left(\eta^{bc}\pa_b\pa_c+m^2\right) A_a=0,\\
\left(\pa_y e^{-2\kappa|y|}\pa_y+m^2\right)f=0,
\end{split}
\end{equation*}for some Kaluza-Klein mass $m$. Furthermore, $f$ has to satisfy the following orbifold boundary conditions \cite{39}:
$$\left.\pa_y f(y)\right|_{y=0, \pi R}=0.$$The solutions are given by \cite{39}:
$$f_n(y)=e^{\kappa|y|} \left[J_1\left(\frac{m_n}{\kappa}e^{\kappa|y|}\right)+b_n Y_{1}\left(\frac{m_n}{\kappa}e^{\kappa|y|}\right)\right]
,$$where $J_{\nu}(z)$ and $Y_{\nu}(z)$ are Bessel functions of first and second kind, $0<m_1<m_2<m_3<\ldots$ are positive solutions of
\begin{equation}\label{eq7_28_9}\begin{split}\left[J_1\left(\frac{z}{\kappa} \right)+\frac{z}{\kappa}J_1'\left(\frac{z}{\kappa} \right) \right]\left[Y_1\left(\frac{ze^{\pi\kappa R}}{\kappa} \right)+\frac{z e^{\pi\kappa R}}{\kappa}Y_1'\left(\frac{ze^{\pi \kappa R}}{\kappa} \right)\right]&\\-\left[Y_1\left(\frac{z}{\kappa} \right)+\frac{z}{\kappa}Y_1'\left(\frac{z}{\kappa} \right) \right]\left[J_1\left(\frac{ze^{\pi\kappa R}}{\kappa} \right)+\frac{z e^{\pi\kappa R}}{\kappa}J_1'\left(\frac{ze^{\pi \kappa R}}{\kappa} \right)\right]&\\
=\frac{z^2e^{\pi\kappa R}}{\kappa^2}\left(J_0\left(\frac{z}{\kappa} \right)Y_0\left(\frac{ze^{\pi\kappa R}}{\kappa} \right)-Y_0\left(\frac{z}{\kappa} \right)J_0\left(\frac{ze^{\pi\kappa R}}{\kappa} \right)\right)&=0;
\end{split}\end{equation}and
\begin{align*}
b_n=-\frac{J_1\left(\frac{m_n}{\kappa} \right)+\frac{m_n}{\kappa}J_1'\left(\frac{m_n}{\kappa} \right) }{Y_1\left(\frac{m_n}{\kappa} \right)+\frac{m_n}{\kappa}Y_1'\left(\frac{m_n}{\kappa} \right)}.
\end{align*}

Under 5$D$ perfectly conducting boundary conditions, the potential is identically zero inside the plates. In the system I depicted in FIG. \ref{f2}, the fields are confined in the three chambers Ia, Ib and Ic independently. In other words, the field modes are the union of three types of modes, each type is nonzero in one of the chambers and zeros elsewhere. Therefore, the Casimir energy of system I is the sum of the Casimir energies in the three chambers, namely,
\begin{align*}
E_{\text{Cas}}\left(b_l, b_r, L_1\right)=E_{\text{Cas}}^{\text{chamber}}(d_{1a})+E_{\text{Cas}}^{\text{chamber}}(d_{1b})+E_{\text{Cas}}^{\text{chamber}}(d_{1c}),
\end{align*}
where $E_{\text{Cas}}^{\text{chamber}}(d)$ is the Casimir energy   in a chamber of width $d$.

For a chamber extended from $x^1=0$ and $x^1=d$, the boundary condition \eqref{eq8_2_1} reads as
\begin{equation}\label{eq7_28_10}
\left.\pa_{\mu}V_{\nu}-\pa_{\nu}V_{\mu}\right|_{x^1=0,b}=0,\quad\quad \text{for}\;\;\mu,\nu\neq 1.
\end{equation}For the $x^2$ and $x^3$ direction, we can take $L_2, L_3\rightarrow \infty$ from the beginning.
Subjected to the boundary condition \eqref{eq7_28_10}, the eigenmodes of the potential are given by:\\ For the Kaluza-Klein zero mode,
\begin{equation}
\begin{split}\begin{aligned}
V_0^0=&0,\\
V_1^0=&\alpha_1 \cos\frac{\pi k x^1}{d}e^{ik_2x^2+ik_3x^3-i\omega t},\\
V_2^0=&\alpha_2\sin\frac{\pi k x^1}{d}e^{ik_2x^2+ik_3x^3-i\omega t},\\
V_3^0=&\alpha_3\sin\frac{\pi k x^1}{d}e^{ik_2x^2+ik_3x^3-i\omega t},\end{aligned} \hspace{2cm}k=0,1,2,\ldots,
\end{split}
\end{equation}subjected to the condition \begin{equation}\label{eq7_28_12}-\frac{\pi k}{d}\alpha_1+i\alpha_2 k_2+i\alpha_3k_3=0.\end{equation} The dispersion relation is $$\omega^2= \left(\frac{\pi k}{d}\right)^2+k_{\perp}^2,$$ where $k_{\perp}=\sqrt{k_2^2+k_3^2}$. \eqref{eq7_28_12} shows that there are two polarizations in this case. \\For the Kaluza-Klein excitation modes,
\begin{equation*}
\begin{split}\begin{aligned}
V_0^n=&\alpha_0\sin\frac{\pi k x^1}{d}e^{ik_2x^2+ik_3x^3-i\omega t}f_n(y),\\
V_1^n=&\alpha_1 \cos\frac{\pi k x^1}{d}e^{ik_2x^2+ik_3x^3-i\omega t}f_n(y),\\
V_2^n=&\alpha_2\sin\frac{\pi k x^1}{d}e^{ik_2x^2+ik_3x^3-i\omega t}f_n(y),\\
V_3^n=&\alpha_3\sin\frac{\pi k x^1}{d}e^{ik_2x^2+ik_3x^3-i\omega t}f_n(y),\end{aligned} \hspace{2cm}k=0,1,2,\ldots,
\end{split}
\end{equation*}subjected to the condition \begin{equation}\label{eq7_28_11}i\omega\alpha_0-\frac{\pi k}{d}\alpha_1+i\alpha_2 k_2+i\alpha_3k_3=0.\end{equation} The dispersion relation is  $$\omega^2= \left(\frac{\pi k}{d}\right)^2+k_{\perp}^2+m_n^2.$$   \eqref{eq7_28_11} shows that there are three polarizations in this case.

To compute the Casimir energy in a chamber, we use zeta regularization \eqref{eq7_29_1}. Denote by $A=L_2L_3$, the finite temperature zeta function $\zeta_T^{\text{chamber}}(s;d)$ for a chamber of width $d$ is given by
\begin{equation*}
\begin{split}
&\zeta_T^{\text{chamber}}(s;d) \\=&\frac{A}{2\pi}\int_0^{\infty}\left(h_0\sum_{\ell=-\infty}^{\infty}\sum_{k=0}^{\infty}\!'' \left(\left[\frac{\pi k}{d}\right]^2+k_{\perp}^2+[2\pi \ell T]^2\right)^{-s}
+h  \sum_{n=1}^{\infty}\sum_{\ell=-\infty}^{\infty}\sum_{k=0}^{\infty} \left(\left[\frac{\pi k}{d}\right]^2+k_{\perp}^2+m_n^2+[2\pi \ell T]^2\right)^{-s}\right) k_{\perp}dk_{\perp},\end{split}\end{equation*}where $h_0=2$ and $h=3$ are the number of polarizations for the Kaluza-Klein zero mode and Kaluza-Klein excitation modes respectively; and double prime $''$ means that for $\ell=0$, the term $k=0$ is omitted. Integrating out $k_{\perp}$, we find that
\begin{equation*}\begin{split}
\zeta_T^{\text{chamber}}(s;d)=&\frac{A}{4\pi(s-1)}\left(h_0\sum_{\ell=-\infty}^{\infty}\sum_{k=0}^{\infty}\!''\left(\left[\frac{\pi k}{d}\right]^2 +[2\pi \ell T]^2\right)^{1-s}+h
 \sum_{n=1}^{\infty}\sum_{\ell=-\infty}^{\infty} \sum_{k=0}^{\infty}\left(\left[\frac{\pi k}{d}\right]^2 +m_n^2+[2\pi \ell T]^2\right)^{1-s}\right)\\
=&\frac{A}{4\pi(s-1)}\left(h_0 \sum_{k=1}^{\infty} \left( \frac{\pi k}{d} \right)^{2-2s}+h_0\sum_{\ell=1}^{\infty} \left( 2\pi \ell T \right)^{2-2s}
+2h_0\sum_{\ell=1}^{\infty}\sum_{k=0}^{\infty}\!'\left(\left[\frac{\pi k}{d}\right]^2 +[2\pi \ell T]^2\right)^{1-s}\right.\\&\left.
+\frac{h}{2} \sum_{n=1}^{\infty}\sum_{\ell=-\infty}^{\infty}  \left( m_n^2+[2\pi \ell T]^2\right)^{1-s}+h
 \sum_{n=1}^{\infty}\sum_{\ell=-\infty}^{\infty} \sum_{k=0}^{\infty}\!'\left(\left[\frac{\pi k}{d}\right]^2 +m_n^2+[2\pi \ell T]^2\right)^{1-s}\right) \\
=&\mathcal{Y}_0(s)+\frac{Ah_0}{4\pi(s-1)}\left(\frac{\pi}{d}\right)^{2 -2s}\zeta_R(2s-2)+ h_0\mathcal{Z}_1(s;d) +h\sum_{n=1}^{\infty}\mathcal{Z}_2(s,m_n;d),
\end{split}
\end{equation*}   where prime $^{\prime}$ means that the term with $k=0$ is multiplied by a factor one-half; $\zeta_R(s)$ is the Riemann zeta function $\displaystyle \zeta_R(s)=\sum_{n=1}^{\infty}n^{-s}$;
\begin{align*}
\mathcal{Y}_0(s)=\frac{A}{4\pi(s-1)}\left(h_0  \left( 2\pi  T \right)^{2-2s}\zeta_R(2s-2)+\frac{h}{2} \sum_{n=1}^{\infty}\sum_{\ell=-\infty}^{\infty}  \left( m_n^2+[2\pi \ell T]^2\right)^{1-s}\right)
\end{align*}is independent of $d$;
\begin{equation*}
\begin{split}
\mathcal{Z}_1(s;d )
=&\frac{A}{2\pi(s-1)}\sum_{\ell=1}^{\infty}\sum_{k=0}^{\infty}\!'\left(\left[\frac{\pi k}{d}\right]^2+[2\pi \ell T]^2\right)^{1-s};
\end{split}
\end{equation*}and
\begin{equation*}
\begin{split}
&\mathcal{Z}_2(s,m_n;d )=\frac{A}{4\pi(s-1)}\sum_{\ell=-\infty}^{\infty} \sum_{k=0}^{\infty}\!'\left(\left[\frac{\pi k}{d}\right]^2 +m_n^2+[2\pi \ell T]^2\right)^{1-s}.  \end{split}
\end{equation*}Using the formula
$$\sum_{k=0}^{\infty}\!'\exp\left(-t\left[\frac{\pi k}{d}\right]^2\right)=\frac{d}{\sqrt{\pi t}}\sum_{k=0}^{\infty}\!'\exp\left(-\frac{k^2d^2}{t}\right),$$ we find that
\begin{equation*}
\begin{split}
\mathcal{Z}_1(s;d )=&\frac{A}{2\pi\Gamma(s )}\int_0^{\infty}t^{s-2}\sum_{\ell=1}^{\infty}\sum_{k=0}^{\infty}\!'
\exp\left\{-t\left(\left[\frac{\pi k}{d}\right]^2 +[2\pi \ell T]^2\right)\right\}dt \\
=& \frac{Ad}{2\pi^{\frac{3}{2}}\Gamma(s )}\int_0^{\infty}t^{s-\frac{5}{2}}\sum_{\ell=1}^{\infty}\sum_{k=0}^{\infty}\!'
\exp\left\{-t (2\pi \ell T)^2-\frac{k^2d^2}{t}\right\}dt \\
=& \frac{Ad}{4\pi^{\frac{3}{2}}}\frac{\Gamma\left(s-\frac{3}{2}\right)}{\Gamma(s)}(2\pi T)^{3-2s}\zeta_R(2s-3)+\frac{Ad}{\pi^{\frac{3}{2}}\Gamma(s)}\sum_{l=1}^{\infty}\sum_{k=1}^{\infty}\left(\frac{kd}{2\pi \ell T}\right)^{s-\frac{3}{2}}K_{s-\frac{3}{2}}\left(4\pi l k dT\right),
\end{split}
\end{equation*}where $K_{\nu}(z)$ is the modified Bessel function of second kind. Similarly,
\begin{equation*}
\begin{split}
\mathcal{Z}_2(s,m_n;d )=&\frac{Ad}{4\pi^{\frac{3}{2}}}\frac{\Gamma\left(s-\frac{3}{2}\right)}{\Gamma(s)}\sum_{l=0}^{\infty}\!'\left(m_n^2+[2\pi\ell T]^2\right)^{\frac{3}{2}-s}\\&+
\frac{Ad}{\pi^{\frac{3}{2}}\Gamma(s)}\sum_{l=0}^{\infty}\!'\sum_{k=1}^{\infty}\left(\frac{kd}{\sqrt{m_n^2+[2\pi \ell T]^2}}\right)^{s-\frac{3}{2}}K_{s-\frac{3}{2}}\left(2kd \sqrt{m_n^2+[2\pi \ell T]^2}\right).
\end{split}
\end{equation*}Therefore,
\begin{equation}\label{eq7_29_2}
\begin{split}
\zeta_T^{\text{chamber}}(s;d) =&\mathcal{Y}_0(s)+d\mathcal{Y}_1(s)+\frac{Ah_0}{4\pi\Gamma(s)}\left(\frac{\pi}{d}\right)^{2 -2s}\Gamma(s-1)\zeta_R(2s-2)+\frac{Ah_0d}{\pi^{\frac{3}{2}}\Gamma(s)}\sum_{l=1}^{\infty}\sum_{k=1}^{\infty}\left(\frac{kd}{2\pi \ell T}\right)^{s-\frac{3}{2}}K_{s-\frac{3}{2}}\left(4\pi l k dT\right)\\&+\frac{Ahd}{\pi^{\frac{3}{2}}\Gamma(s)}\sum_{n=1}^{\infty}\sum_{l=0}^{\infty}\!'\sum_{k=1}^{\infty}\left(\frac{kd}{\sqrt{m_n^2+[2\pi \ell T]^2}}\right)^{s-\frac{3}{2}}K_{s-\frac{3}{2}}\left(2kd \sqrt{m_n^2+[2\pi \ell T]^2}\right),
\end{split}
\end{equation}where
\begin{equation*}
\begin{split}
\mathcal{Y}_1(s)=\frac{A h_0}{4\pi^{\frac{3}{2}}}\frac{\Gamma\left(s-\frac{3}{2}\right)}{\Gamma(s)}(2\pi T)^{3-2s}\zeta_R(2s-3)+\frac{A h }{4\pi^{\frac{3}{2}}}\frac{\Gamma\left(s-\frac{3}{2}\right)}{\Gamma(s)}\sum_{l=0}^{\infty}\!'\left(m_n^2+[2\pi\ell T]^2\right)^{\frac{3}{2}-s}
\end{split}
\end{equation*}is also independent of $d$. Notice that $1/\Gamma(s)$ has a zero at $s=0$, and $$\Gamma(s-1)\zeta_R(2s-2)=\pi^{ 2s-\frac{5}{2}}\Gamma\left(\frac{3}{2}-s\right)\zeta_R(3-2s)$$ only has poles at $s=1$ and $s=3/2$. Therefore, the nontrivial contribution to $\zeta_T^{\text{chamber}}(0;d) $ \eqref{eq7_29_2} only comes from the first two terms, i.e.,
$$\zeta_T^{\text{chamber}}(0;d) = \mathcal{Y}_0(0)+d\mathcal{Y}_1(0),$$which depends on $d$ linearly. Moreover,
\begin{equation}\label{eq7_29_3}
\begin{split}
\zeta_T^{\text{chamber}\prime} (0;d) =&\mathcal{Y}_0'(0)+d\mathcal{Y}_1'(0)+\frac{Ah_0}{8\pi } \frac{\zeta_R(3)}{d^2} +\frac{2\sqrt{2}Ah_0T^{\frac{3}{2}}}{\sqrt{d} }\sum_{l=1}^{\infty}\sum_{k=1}^{\infty}\left(\frac{ \ell  }{k }\right)^{ \frac{3}{2}}K_{ \frac{3}{2}}\left(4\pi l k dT\right)\\&+\frac{Ah }{\pi^{\frac{3}{2}} \sqrt{d} }\sum_{n=1}^{\infty}\sum_{l=0}^{\infty}\!'\sum_{k=1}^{\infty}\left(\frac{\sqrt{m_n^2+[2\pi \ell T]^2}}{k }\right)^{ \frac{3}{2}}K_{ \frac{3}{2}}\left(2kd \sqrt{m_n^2+[2\pi \ell T]^2}\right).
\end{split}
\end{equation}The first two terms are linear in $d$ and the other terms go to zero when $d\rightarrow \infty$. The finite temperature Casimir energy in the chamber of width $d$ is then given by
\begin{equation*}
E_{\text{Cas}}^{\text{chamber}}(d)=-\frac{T}{2}\left(\zeta_T^{\text{chamber}\prime} (0) +[\log \mu^2]\zeta_T^{\text{chamber}}(0)\right).
\end{equation*}Applying the renormalization scheme described in Section \ref{s1}, we find that the renormalized Casimir energy of the parallel plate system with $5D$ induced perfectly conducting condition is given by
\begin{equation} \label{eq7_29_5}
E_{\text{Cas}, 5D}^{\parallel}=\lim_{\substack{L_1,b_l,b_r, L_2,L_3\rightarrow\infty\\ \eta_l,\eta_r, a=b_r-b_l\;\text{fixed}}} \Bigl\{E_{\text{Cas}}^{\text{chamber}}(d_{1a})+
E_{\text{Cas}}^{\text{chamber}}(d_{1b})+E_{\text{Cas}}^{\text{chamber}}(d_{1c})-E_{\text{Cas}}^{\text{chamber}}(d_{2a})-E_{\text{Cas}}^{\text{chamber}}(d_{2b})
-E_{\text{Cas}}^{\text{chamber}}(d_{2c})\Bigr\}.
\end{equation} The terms that are linear in $d$ in $E_{\text{Cas}}^{\text{chamber}}(d )$ will cancel since $d_{1a}+d_{1b}+d_{1c}=d_{2a}+d_{2b}+d_{2c}=L_1-t_l-t_r$. The other terms in $E_{\text{Cas}}^{\text{chamber}}(d )$ goes to zero when $d\rightarrow \infty$.
Since the limits $L_1,b_l,b_r \rightarrow\infty$ with $\eta_l, \eta_r$ and $a$ fixed is the same as letting  $d_{1a}, d_{1c}, d_{2a}, d_{2b}, d_{2c}$ go to infinity but keeping $d_{1b}=a$ fixed, only the part of $E_{\text{Cas}}^{\text{chamber}}(d_{1b} )$ that goes to zero when $d_{1b}\rightarrow \infty$ remains in the renormalized Casimir energy \eqref{eq7_29_5}, which gives
\begin{equation}\label{eq7_30_3}\begin{split}
E_{\text{Cas},5D}^{\parallel}=&E_{\text{Cas},0}^{\parallel}+hE_{\text{Cas},1}^{\parallel},
\end{split}
\end{equation}where
\begin{equation}\begin{split}
E_{\text{Cas},0}^{\parallel}=-\frac{AT}{8\pi } \frac{\zeta_R(3)}{a^2}-\frac{2\sqrt{2}AT^{\frac{5}{2}}}{\sqrt{a} }\sum_{l=1}^{\infty}\sum_{k=1}^{\infty}\left(\frac{ \ell  }{k }\right)^{ \frac{3}{2}}K_{ \frac{3}{2}}\left(4\pi l k aT\right),\end{split}\end{equation}and \begin{equation}\begin{split}&E_{\text{Cas},1}^{\parallel}=-\frac{AT }{2\pi^{\frac{3}{2}} \sqrt{a}}\sum_{n=1}^{\infty}\sum_{l=0}^{\infty}\!'\sum_{k=1}^{\infty}\left(\frac{\sqrt{m_n^2+[2\pi \ell T]^2}}{k }\right)^{ \frac{3}{2}}K_{ \frac{3}{2}}\left(2ka \sqrt{m_n^2+[2\pi \ell T]^2}\right).
\end{split}\end{equation}Since $K_{\nu}(z)$ is positive for positive $z$ and $\zeta_R(z)$ is positive for $z>1$, the Casimir energy is always negative. The term $ E_{\text{Cas},0}^{\parallel}$ is the finite temperature Casimir energy for a pair of parallel perfectly conducting plates in 4$D$ Minkowski spacetime.
The finite temperature Casimir force acting on the parallel plates with $5D$ induced perfectly conducting condition  is given by
\begin{equation}\label{eq7_30_4}
\begin{split}
F_{\text{Cas},5D}^{\parallel}=&-\frac{\pa E_{\text{Cas}}^{\parallel}}{\pa a}= F_{\text{Cas},0}^{\parallel}+hF_{\text{Cas},1}^{\parallel},
\end{split}
\end{equation}where
\begin{equation}\begin{split}F_{\text{Cas},0}^{\parallel}=&-\frac{A T}{4\pi } \frac{\zeta_R(3)}{a^3}
-\frac{4\sqrt{2}A T^{\frac{5}{2}}}{a^{\frac{3}{2}}}\sum_{l=1}^{\infty}\sum_{k=1}^{\infty}\left(\frac{ \ell  }{k }\right)^{ \frac{3}{2}}K_{ \frac{3}{2}}\left(4\pi l k aT\right)-\frac{8\pi\sqrt{2}A T^{\frac{7}{2}}}{\sqrt{a} }\sum_{l=1}^{\infty}\sum_{k=1}^{\infty} \frac{ \ell^{\frac{5}{2}}  }{k^{\frac{1}{2}} } K_{ \frac{1}{2}}\left(4\pi l k aT\right),\\
F_{\text{Cas},1}^{\parallel}=&- \frac{AT }{ \pi^{\frac{3}{2}} a^{\frac{3}{2}}}\sum_{n=1}^{\infty}\sum_{l=0}^{\infty}\!'\sum_{k=1}^{\infty}\left(\frac{\sqrt{m_n^2+[2\pi \ell T]^2}}{k }\right)^{ \frac{3}{2}}K_{ \frac{3}{2}}\left(2ka \sqrt{m_n^2+[2\pi \ell T]^2}\right)\\
&-\frac{  AT }{ \pi^{\frac{3}{2}} \sqrt{a}}\sum_{n=1}^{\infty}\sum_{l=0}^{\infty}\!'\sum_{k=1}^{\infty}\frac{\left(\sqrt{m_n^2+[2\pi \ell T]^2}\right)^{ \frac{5}{2}}}{k^{\frac{1}{2}} }K_{ \frac{1}{2}}\left(2ka \sqrt{m_n^2+[2\pi \ell T]^2}\right).
\end{split}
\end{equation}The term $ F_{\text{Cas},0}^{\parallel}$ is the finite temperature Casimir force acting on a pair of $4D$ perfectly conducting plates \cite{50}. The term $hF_{\text{Cas},1}^{\parallel}$ is the correction due to extra dimension. Both $F_{\text{Cas},0}^{\parallel}$ and $F_{\text{Cas},1}^{\parallel}$ are always negative. Therefore, the Casimir force due to $5D$ perfectly conducting condition is always attractive, and have larger magnitude than the Casimir force in the $4D$ Minkowski spacetime. It is also easy to verify that the Casimir force is a decreasing function of the plate separation $a$.

By taking the limit $T\rightarrow 0$ of \eqref{eq7_30_3}  using
\begin{equation}\label{eq8_4_5}T\sum_{\ell=0}^{\infty}\!'f(2\pi\ell T) \quad\xrightarrow{T\rightarrow 0}\quad\frac{1}{2\pi}\int_0^{\infty} f(\xi)d\xi,\end{equation} and
\begin{equation*}\begin{split}
\int_0^{\infty}\xi^{\frac{3}{2}}K_{\frac{3}{2}}(2\alpha\xi)=&\frac{\sqrt{ \pi}}{4\alpha^{\frac{5}{2}}},\\
\int_0^{\infty}\left(\xi^2+m^2\right)^{\frac{3}{4}}K_{\frac{3}{2}}\left(2\alpha\sqrt{\xi^2+m^2}\right)d\xi=&\frac{\alpha^{\frac{3}{2}}}{2}
\int_0^{\infty}\int_0^{\infty} t^{\frac{1}{2}}\exp\left(-\frac{\xi^2+m^2}{t}-t\alpha^2\right)dtd\xi\\
=&\frac{\sqrt{\pi}\alpha^{\frac{3}{2}}}{4}
 \int_0^{\infty} t \exp\left(-\frac{ m^2}{t}-t\alpha^2\right)dt \\=&\frac{\sqrt{\pi}m^2}{2\alpha^{\frac{1}{2}}}K_2\left(2\alpha m\right),
\end{split}
\end{equation*}we find that the zero temperature Casimir energy under $5D$ induced perfectly conducting condition  is given  by
\begin{equation}\label{eq8_4_7}\begin{split}
E_{\text{Cas},5D}^{\parallel, T=0}=& -\frac{ Ah_0\pi^2}{ 1440a^3}-\frac{Ah  }{8\pi^2a}\sum_{n=1}^{\infty} \sum_{k=1}^{\infty}\left(\frac{ m_n }{k }\right)^{2}K_{ 2}\left(2ka m_n \right).
\end{split}\end{equation}It follows that the zero temperature Casimir force under $5D$ induced perfectly conducting condition is equal to
\begin{equation}\label{eq8_4_8}\begin{split}
F_{\text{Cas},5D}^{\parallel, T=0}=& -\frac{ Ah_0\pi^2}{ 480a^4}-\frac{3Ah  }{8\pi^2a^2}\sum_{n=1}^{\infty} \sum_{k=1}^{\infty}\left(\frac{ m_n }{k }\right)^{2}K_{ 2}\left(2ka m_n \right)-\frac{Ah  }{4\pi^2a}\sum_{n=1}^{\infty} \sum_{k=1}^{\infty} \frac{ m_n^3 }{k } K_{ 1}\left(2ka m_n \right).
\end{split}\end{equation}

Under $5D$ induced perfectly conducting conditions, the zero temperature Casimir energy \eqref{eq8_4_7} and zero temperature Casimir force \eqref{eq8_4_8} for electromagnetic field are similar to the results obtained in \cite{14}. However, one should note that the equation  determining the Kaluza-Klein masses $m_n$ for electromagnetic field \eqref{eq7_28_9} is different from the equation determining the Kaluza-Klein masses for scalar field \cite{39}, and they satisfy different asymptotic behaviors.

\section{Casimir effect with 4D perfectly conducting conditions}\label{s4}
In this section, we treat the two parallel perfectly conducting plates as $4D$ plates.
Instead of the action in vacuum \eqref{eq8_2_2}, we incorporate a conserve current $K^{\mu}$ and the $5D$ bulk action becomes
\begin{equation*}
S=-\int d^4x  \int dy \sqrt{|g|}  \left(\frac{1}{4}F_{\mu\nu}F^{\mu\nu}+K^{\mu}A_{\mu}\right).
\end{equation*}The equation of motion is then given by
\begin{equation} \label{eq8_2_3}
  \sqrt{|g|}^{\;-1} \pa_{\mu}\left( \sqrt{|g|}F^{\mu\nu}\right) = \sqrt{|g|}^{\;-1} \pa_{\mu} \sqrt{|g|}g^{\mu\kappa}g^{\nu\eta}\left( \pa_{\kappa}V_{\eta}- \pa_{\eta} V_{\kappa} \right)  =K^{\nu}.
\end{equation}Conservation law for $K^{\mu}$ implies that
$$\sqrt{|g|}^{\;-1} \pa_{\mu}\left( \sqrt{|g|}K^{\mu }\right) =0.$$
It is understood that $K^{\mu}=0$ in vacuum. Since the plates are $4D$ objects, we also impose the condition $K^y=0$. Then as in the previous section, we can impose the Lorentz gauge and the axial gauge $V_y=0$. Writing $V_a$ and $K_a$ in the form $V_a(x,y)=A_a(x)f(y)$ and $K_a(x,y)=J_a(x)f(y)e^{2\kappa|y|}$, we find as in previous section that Kaluza-Klein decomposition gives rise to a tower of fields $A_a^ndx^a$, $n=0,1,2,\ldots$, satisfying
\begin{equation}\label{eq7_30_1}\left(\eta^{bc}\pa_b\pa_c+m_n^2\right) A_a^n=\left(\pa_0^2-\pa_1^2-\pa_2^2-\pa_3^2+m_n^2\right)A_a^n=J_a^n,\end{equation}   subject to the Lorentz condition \begin{equation}\label{eq7_30_2}\pa_a\left(A^n\right)^a=\pa_0A_0^n-\pa_1A_1^n-\pa_2A_2^n-\pa_3A_3^n=0\end{equation}and the conservation law
\begin{equation}\label{eq8_2_5}\pa_a\left(J^n\right)^a=\pa_0J_0^n-\pa_1J_1^n-\pa_2J_2^n-\pa_3J_3^n=0.\end{equation}
By convention, $n=0$ refers to the zero mode and $m_0=0$.

Let us forget the superscript $n$ for the moment. Denote by $\phi$ and $\mathbf{A}=(A^1,A^2,A^3)$   the potentials with $\phi=A_0$, $A^1=-A_1, A^2=-A_2, A^3=-A_3$. Let $\mathbf{E}=(E^1,E^2,E^3)$ and $\mathbf{B}=(B^1,B^2,B^3)$  be the electric and magnetic fields with $$\mathbf{B}=\nabla\times\mathbf{A},\hspace{1cm}\mathbf{E}=-\frac{\pa\mathbf{A}}{\pa t}-\nabla\phi.$$ Then\begin{align*}E^1=&\pa_0A_1-\pa_1A_0, \quad E^2=\pa_0A_2-\pa_2A_0, \quad E^3=\pa_0A_3-\pa_3A_0,\\ B^1=&\pa_3A_2-\pa_2A_3,\quad B^2=\pa_1A_3-\pa_3A_1, \quad B^3=\pa_2A_1-\pa_1A_2;\end{align*} and one obtains immediately two of the Maxwell's equations:
$$\nabla\cdot\mathbf{B}= 0,\quad \nabla\times \mathbf{E}+\frac{\pa \mathbf{B}}{\pa t}=0.$$  Let $\rho$ and $\mathbf{J}=(J^1,J^2,J^3)$ be the free charges and free currents with $\rho=J_0, J^1=-J_1, J^2=-J_2$ and $J^3=-J_3$.   The equation of motion \eqref{eq7_30_1} and the Lorentz condition
\eqref{eq7_30_2} imply the equations
\begin{equation}
\begin{split}
\nabla\cdot\mathbf{E}+m^2\phi=\rho,\quad
\nabla\times\mathbf{B}-\frac{\pa\mathbf{E}}{\pa t}+m^2\mathbf{A}=\mathbf{J},
\end{split}
\end{equation}which are generalizations of the remaining two Maxwell's equations by Proca \cite{40}, called Proca equations. The Casimir effect on perfectly conducting plates due to Proca field (massive vector field) has been studied by Barton and Dombey \cite{41,42} (see also our work \cite{43}).
As mentioned before, $\rho=0$ and $\mathbf{J} =\mathbf{0}$ in vacuum. For the plates, assume that they are ohmic conductors, i.e.,
$$\mathbf{J}=\sigma \mathbf{E},$$ where $\sigma$ is the conductivity. Perfectly conducting condition amounts to taking the limit $\sigma\rightarrow\infty$. With $\mathbf{J}=\sigma \mathbf{E}$, we have
\begin{align*}J_1=&\sigma\left(\pa_1A_0-\pa_0A_1\right), \quad J_2=\sigma\left(\pa_2A_0-\pa_0A_2\right), \quad J_3=\sigma\left(\pa_3A_0-\pa_0A_3\right);\end{align*}
and the component $J_0$ is determined from the continuity equation \eqref{eq8_2_5}. Under the assumption $A_a(x)=\hat{A}_a(\mathbf{x})e^{-i\omega t}$, $J_a(x)=\hat{J}_a(\mathbf{x})e^{-i\omega t}$, where $\mathbf{x}=(x^1,x^2,x^3)$, we find that
$$J_0=-\frac{i\sigma}{\omega}\left(\pa_0^2-\pa_1^2-\pa_2^2-\pa_3^2\right)A_0.$$The equation of motion can then be rewritten as
\begin{equation*}\begin{split}\left(1+\frac{i\sigma}{\omega}\right)\left(\pa_0^2-\pa_1^2-\pa_2^2-\pa_3^2\right)A_0^n+m_n^2A_0^n=&0,\\
\left(\pa_0^2+\sigma\pa_0-\pa_1^2-\pa_2^2-\pa_3^2\right)A_j^n+m_n^2A_j^n=&\sigma\pa_j A_0^n,\quad j=1,2,3.
\end{split}
\end{equation*}
For nonzero $\sigma$, the plane waves solutions $\displaystyle A_a^n=c_ae^{ik_1x^1+ik_2x^2+ik_3x^3}$ of these equations can be divided into transverse waves with $\nabla\cdot\mathbf{A}=-(\pa_1A_1+\pa_2A_2+\pa_3A_3)=0$ and $A_0=0$ and longitudinal waves with $\nabla\times \mathbf{A}=0$. The dispersion relations for the transverse waves and longitudinal waves are given respectively by \begin{align}\label{eq8_2_6}\omega^2\left(1+\frac{i\sigma}{\omega}\right)=k_1^2+k_{\perp}^2+m_n^2\end{align}
and $$\omega^2=k_1^2+k_{\perp}^2+\frac{m^2}{ 1+\frac{i\sigma}{\omega}}.$$In the perfect conductor limit $(\sigma\rightarrow \infty)$, it follows from \eqref{eq8_2_6} that the eigenfrequency $\omega$ of the transverse modes has to be zero, i.e., no transverse modes can exist in perfectly conducting materials. However, longitudinal modes with $$ \omega^2=k_1^2+k_{\perp}^2$$can exist in perfectly conducting objects.

Returning to the system of parallel perfectly conducting plates. For the boundary conditions, one requires the electric field and magnetic field to vanish inside the perfectly conducting plates. However, in the massive sector, the potentials $\phi$ and $\mathbf{A}$ or equivalently, the one form $A_adx^a$, \emph{do  not} have to vanish inside the plates. On the other hand, one also need to impose the conditions that \emph{all the components of $A_adx^a$ being continuous on the interfaces} \cite{41,42,51}.  The Lorentz condition \eqref{eq7_30_2} then implies that $\pa_1A_1$ is also continuous. The eigenmodes can then be divided into two types:  the discrete modes and the continuum mode.  The discrete modes have two polarizations called type 1 and type 2 discrete modes. They are those modes where $A_{a}dx^a$ vanishes   inside the perfectly conducting plates. When $m\neq 0$, there is a nontrivial  continuum mode   where $A_adx^a$ does not vanish inside the plates.

For type 1 or type 2 discrete modes, since $A_adx^a$ vanishes identically in the perfectly conducting plates, the modes in the three chambers of system I in FIG. \ref{f2} are independent. Therefore, as in the previous section, the contribution to the Casimir energy of system I from the discrete modes is the sum of the contributions from the three chambers. In a chamber extended from $x^1=0$ to $x^1=d$, the discrete modes of type I are given by
\begin{equation*}
\begin{split}\begin{aligned}
A_0^{n,1}=&0,\\
A_1^{n,1}=&0,\\
A_2^{n,1}=&-k_3\sin\frac{\pi k x^1}{d}e^{ik_2x^2+ik_3x^3-i\omega t},\\
A_3^{n,1}=&k_2\sin\frac{\pi k x^1}{d}e^{ik_2x^2+ik_3x^3-i\omega t},\end{aligned} \hspace{2cm}k=1,2,\ldots;
\end{split}
\end{equation*}and the discrete modes of type II are given by
\begin{equation*}
\begin{split}\begin{aligned}
A_0^{n,2}=&-\frac{k_{\perp}^2}{\omega}\sin\frac{\pi k x^1}{d}e^{ik_2x^2+ik_3x^3-i\omega t},\\
A_1^{n,2}=&0,\\
A_2^{n,2}=&k_2\sin\frac{\pi k x^1}{d}e^{ik_2x^2+ik_3x^3-i\omega t},\\
A_3^{n,2}=&k_3\sin\frac{\pi k x^1}{d}e^{ik_2x^2+ik_3x^3-i\omega t},\end{aligned} \hspace{2cm}k= 1,2,\ldots.
\end{split}
\end{equation*}
For both types of discrete modes, $$\omega^2=\left(\frac{\pi k}{d}\right)^2+k_{\perp}^2+m_n^2.$$ The discrete mode contribution to the Casimir energy of the parallel plate system can be computed as in the previous section. In fact,   the only differences now are that: (a) we have two discrete polarizations for both massless and massive sectors, compared to two discrete polarizations for massless sector and three discrete polarizations for massive sector in the previous section; and (b) $k$ starts from one instead of zero. However, the modes corresponding to $k=0$ does not depend on $d$. Therefore, they will be canceled out after subtracting the corresponding Casimir energy of the reference system (system II). As a result, it is easy to see that the contribution to the renormalized Casimir energy of the parallel plate system from the discrete modes is given by \eqref{eq7_30_3} with the replacement $h=3\rightarrow h_0=2$, i.e.,
\begin{equation}\label{eq7_30_5}\begin{split}
E_{\text{Cas}}^{\parallel, \text{discrete}}=&E_{\text{Cas},0}^{\parallel}+h_0E_{\text{Cas},1}^{\parallel},
\end{split}\end{equation}which is again always negative. The discrete mode contribution to the Casimir force is given  by \eqref{eq7_30_4} with the replacement $h\rightarrow h_0$, namely,
 $$F_{\text{Cas}}^{\parallel, \text{discrete}}=F_{\text{Cas},0}^{\parallel}+h_0F_{\text{Cas},1}^{\parallel},$$which is again always attractive. The discrete mode contribution to the zero temperature Casimir energy and zero temperature Casimir force are given by \eqref{eq8_4_7} and \eqref{eq8_4_8} respectively with the replacement $h\rightarrow h_0$.

For the continuum modes, $A_{a}dx^a$ can have longitudinal modes inside the plates. Therefore, $A_{a}dx^a$ are not independent in each chamber of the system I and in the plates. In the chambers Ia, Ib, Ic, the modes can be written as
\begin{equation*}
\begin{split}\begin{aligned}
A_0^{n,3}=&-\omega p_n\left(C_j^ne^{ip_nx^1}+D_j^ne^{-ip_nx^1}\right)e^{ik_2x^2+ik_3x^3-i\omega t},\\
A_1^{n,3}=& p_0^2\left(C_j^ne^{ip_nx^1}-D_j^ne^{-ip_nx^1}\right)e^{ik_2x^2+ik_3x^3-i\omega t},\\
A_2^{n,3}=&p_nk_2\left(C_j^ne^{ip_nx^1}+D_j^ne^{-ip_nx^1}\right)e^{ik_2x^2+ik_3x^3-i\omega t},\\
A_3^{n,3}=&p_nk_3\left(C_j^ne^{ip_nx^1}+D_j^ne^{-ip_nx^1}\right)e^{ik_2x^2+ik_3x^3-i\omega t},\end{aligned}
\end{split}
\end{equation*}where $j=1,3,5$ for $x$ in Ia, Ib, Ic respectively; and $$\omega^2=p_0^2+k_{\perp}^2=p_n^2+k_{\perp}^2+m_n^2.$$ In the left and right plates,
\begin{equation*}
\begin{split}\begin{aligned}
A_0^{n,3}=&-\omega\left(C_j^ne^{ip_0x^1}+D_j^ne^{-ip_0x^1}\right)e^{ik_2x^2+ik_3x^3-i\omega t},\\
A_1^{n,3}=& p_0\left(C_j^ne^{ip_0x^1}-D_j^ne^{-ip_0x^1}\right)e^{ik_2x^2+ik_3x^3-i\omega t},\\
A_2^{n,3}=& k_2\left(C_j^ne^{ip_0x^1}+D_j^ne^{-ip_0x^1}\right)e^{ik_2x^2+ik_3x^3-i\omega t},\\
A_3^{n,3}=& k_3\left(C_j^ne^{ip_0x^1}+D_j^ne^{-ip_0x^1}\right)e^{ik_2x^2+ik_3x^3-i\omega t},\end{aligned}
\end{split}
\end{equation*}where $j=2,4$ for $x$ in the left and right plates respectively. It can be readily check that the electric and magnetic fields $\mathbf{E}$ and $\mathbf{B}$ indeed vanish inside the plates. For notational convenience,  let us denote by $ a_1=b_l-t_l, a_2=b_l, a_3=b_r, a_4=b_r+t_r$ the $x^1$-coordinates of the boundaries of the plates, and $a_0=0, a_5=L_1$ the $x^1$-coordinates of the left end and right end of the system. The continuities of $A_0^n, A_1^n, A_2^n, A_3^n$ imply the following equations:
\begin{equation}\label{eq8_2_7}
\begin{split}
\left\{\begin{aligned}p_n\left(C_1^ne^{ip_na_1}+D_1^ne^{-ip_na_1}\right)=C_2^ne^{ip_0a_1}+D_2^ne^{-ip_0a_1}\\
p_0\left(C_1^ne^{ip_na_1}-D_1^ne^{-ip_na_1}\right)=C_2^ne^{ip_0a_1}-D_2^ne^{-ip_0a_1}\end{aligned}\right.,\quad \left\{\begin{aligned}p_n\left(C_3^ne^{ip_na_2}+D_3^ne^{-ip_na_2}\right)=C_2^ne^{ip_0a_2}+D_2^ne^{-ip_0a_2}\\
p_0\left(C_3^ne^{ip_na_2}-D_3^ne^{-ip_na_2}\right)=C_2^ne^{ip_0a_2}-D_2^ne^{-ip_0a_2}\end{aligned}\right.,\\
\left\{\begin{aligned}p_n\left(C_3^ne^{ip_na_3}+D_3^ne^{-ip_na_3}\right)=C_4^ne^{ip_0a_3}+D_4^ne^{-ip_0a_3}\\
p_0\left(C_3^ne^{ip_na_3}-D_3^ne^{-ip_na_3}\right)=C_4^ne^{ip_0a_3}-D_4^ne^{-ip_0a_3}\end{aligned}\right.,\quad \left\{\begin{aligned}p_n\left(C_5^ne^{ip_na_4}+D_5^ne^{-ip_na_4}\right)=C_4^ne^{ip_0a_4}+D_4^ne^{-ip_0a_4}\\
p_0\left(C_5^ne^{ip_na_4}-D_5^ne^{-ip_na_4}\right)=C_4^ne^{ip_0a_4}-D_4^ne^{-ip_0a_4}\end{aligned}\right..
\end{split}
\end{equation}  We also need to impose some boundary conditions on the artificial boundaries at $x^1=a_0$ and $x^1=a_5$. We can impose the conditions $E^2$ and $E^3$ vanish on these boundaries, which give
\begin{equation}\label{eq8_2_8}
C_1^ne^{ip_na_0}+D_1^ne^{-ip_na_0}=0,\hspace{1cm}C_5^ne^{ip_na_5}+D_5^ne^{-ip_na_5}=0.
\end{equation}\eqref{eq8_2_7} and \eqref{eq8_2_8} give rise to altogether ten linear equations of the ten unknowns $C_1^n, D_1^n,\ldots, C_5^n, D_5^n$ which can be written in a matrix form $N_n(C_1^n\; D_1^n \;\ldots\; C_5^n\; D_5^n)^T=(0\;\ldots\;0)^T$, where $N_n$ is a ten by ten matrix. The eigenfrequencies $\omega$ of continuum modes are those $\omega$ that give rise to nontrivial solutions of $(C_1^n\; D_1^n \;\ldots\; C_5^n\; D_5^n)^T$. Therefore, they are solutions of $\det N_n=0$. Similar computations as in \cite{43} show that this is equivalent to
\begin{equation*}
\begin{split}
F_n(\omega;k_{\perp})=&\left[\left(\left(p_n+p_0\right)^2e^{-ip_0t_l}-\left(p_n-p_0\right)^2e^{ip_0t_l}\right)e^{-ip_nd_{1a}}-\left(p_n^2-p_0^2\right)
\left( e^{-ip_0t_l}- e^{ip_0t_l}\right)e^{ip_nd_{1a}}\right]\\
&\times  \left[\left(\left(p_n+p_0\right)^2e^{-ip_0t_r}-\left(p_n-p_0\right)^2e^{ip_0t_r}\right)e^{-ip_nd_{1c}}-\left(p_n^2-p_0^2\right)
\left( e^{-ip_0t_r}- e^{ip_0t_r}\right)e^{ip_nd_{1c}}\right]e^{-ip_nd_{1b}}\\
&- \left[\left(\left(p_n+p_0\right)^2e^{ip_0t_l}-\left(p_n-p_0\right)^2e^{-ip_0t_l}\right)e^{ip_nd_{1a}}-\left(p_n^2-p_0^2\right)
\left( e^{ip_0t_l}- e^{-ip_0t_l}\right)e^{-ip_nd_{1a}}\right]\\
&\times  \left[\left(\left(p_n+p_0\right)^2e^{ip_0t_r}-\left(p_n-p_0\right)^2e^{-ip_0t_r}\right)e^{ip_nd_{1c}}-\left(p_n^2-p_0^2\right)
\left( e^{ip_0t_r}- e^{-ip_0t_r}\right)e^{-ip_nd_{1c}}\right]e^{ip_nd_{1b}}=0,
\end{split}
\end{equation*}where $\displaystyle p_n=p_n(\omega,k_{\perp})=\sqrt{\omega^2-k_{\perp}^2-m_n^2}.$ When $n=0$, $F_n(\omega, k_{\perp})$ reduces to
 $$F_0(\omega,k_{\perp})=16p_0^4\left(e^{-ip_0L_1}-e^{ip_0L_1}\right).$$ The finite temperature  zeta function $\zeta_T^{\text{cont, I}}(s)$ of continuum modes is given by
\begin{align*}
\zeta_T^{\text{cont, I}}(s)=\frac{A}{ \pi}\sum_{n=0}^{\infty}\sum_{\ell=0}^{\infty}\!'\int_0^{\infty}\sum_{\omega >0} \left(\omega^2+[2\pi\ell T]^2\right)^{-s}\text{Res}_{\omega} \frac{d}{d\omega}\ln F_n(\omega;k_{\perp})k_{\perp}dk_{\perp}.
\end{align*}
To evaluate the sum over residues under the integral sign, we need  the generalized Abel-Plana summation formula \cite{44,45,46} which states that: If $f_0(z), f_+(z)$ and $f_-(z) $ are meromorphic functions that does not have poles on the real and imaginary axes, and
\begin{equation}\label{eq6_24_2}\begin{split}
\lim_{Y\rightarrow \infty} \int_{0}^{\infty}\Bigl\{ f_0(x+iY)- f_+(x+ iY)\Bigr\}dx=0,\hspace{1cm}\lim_{X\rightarrow \infty} \int_{0}^{\infty}\Bigl\{ f_0(X+iy)- f_+(X+ iy)\Bigr\}dy=0,\\
\lim_{Y\rightarrow \infty} \int_0^{\infty} \Bigl\{f_0(x-iY)- f_-(x- iY)\Bigr\}dx=0,\hspace{1cm}\lim_{X\rightarrow \infty} \int_0^{\infty} \Bigl\{f_0(X-iy)- f_-(X- iy)\Bigr\}dy=0,
\end{split}
\end{equation} then
\begin{equation}\label{eq6_24_1}
\begin{split}
&\sum_{  \text{Re}\; z\geq 0}  \text{Res}_{z}f_0(z) -\sum_{\substack{ \text{Re}\; z\geq 0\\ \text{Im}\;z\geq 0}}  \text{Res}_{z}f_+(z)
 -\sum_{\substack{ \text{Re}\; z\geq 0\\ \text{Im}\;z\leq 0}}   \text{Res}_{z}f_-(z) \\=& - \frac{1}{2\pi }\int_0^{\infty} \Bigl\{
 f_0(iy)-f_+(iy)\Bigr\}dy- \frac{1}{2\pi }\int_0^{\infty} \Bigl\{
 f_0(-iy)-f_-(-iy)\Bigr\}dy-\frac{1}{2\pi i}\int_0^{\infty} \Bigl\{f_+(x)-f_-(x)\Bigr\}dx.
\end{split}
\end{equation} Setting
\begin{equation*}
\begin{split}
f_{n,0}(z, k_{\perp})=&\sum_{\ell=0}^{\infty}\!'\left(z^2+[2\pi\ell T]^2\right)^{-s} \frac{d}{dz}\ln F_n\left(z,k_{\perp}\right), \\ f_{n,\pm}(z, k_{\perp})=&\sum_{\ell=0}^{\infty}\!'\left(z^2+[2\pi\ell T]^2\right)^{-s} \frac{d}{dz}\ln\left\{(p_n+p_0)^4e^{\mp ip_n (d_{1a}+d_{1b}+d_{1c})\mp ip_0(t_l+t_r)}\right\},
\end{split}
\end{equation*}
it is easy to verify that the conditions in \eqref{eq6_24_2} are satisfied. Therefore, \eqref{eq6_24_1} implies that
\begin{equation*}
\begin{split}
\zeta_T^{\text{cont, I}}(s)=&\Lambda_{T,1}^{\text{I}}(s)+\Lambda_{T,2}^{\text{I}}(s),
\end{split}
\end{equation*}where
\begin{equation*}
\begin{split}
\Lambda_{T,1}^{\text{I}}(s)=\frac{A}{\pi^2}\sum_{n=0}^{\infty}\sum_{\ell=0}^{\infty}\!'\int_0^{\infty}\left\{(d_{1a}+d_{1b}+d_{1c})\int_{\sqrt{k_{\perp}^2+m_n^2}}^{\infty}
\frac{\left(x^2+[2\pi \ell T]^2\right)^{-s}}{\sqrt{x^2-k_{\perp}^2-m_n^2}}dx+(t_l+t_r)\int_{ k_{\perp} }^{\infty}
\frac{\left(x^2+[2\pi \ell T]^2\right)^{-s}}{\sqrt{x^2-k_{\perp}^2} }dx\right\}k_{\perp}dk_{\perp},\end{split}
\end{equation*}and\begin{equation}\label{eq8_5_4}
\begin{split}\Lambda_{T,2}^{\text{I}}(s)=&\frac{A\sin\pi s}{ \pi^2}\sum_{n=0}^{\infty}\sum_{\ell=0}^{\infty}\!'\int_0^{\infty}\int_{2\pi\ell T}^{\infty}\left(\xi^2-[2\pi\ell T]^2\right)^{-s}\frac{d}{d\xi}\ln \Xi^{\text{I}}(\xi,k_{\perp})d\xi k_{\perp}dk_{\perp},\\
\Xi^{\text{I}}(\xi,k_{\perp})= &
\left[\left(1-\Delta_n^2e^{-2q_0t_l}\right)-\Delta_n \left(1-e^{-2q_0t_l}\right)e^{-2q_n d_{1a}}\right]\left[\left(1-\Delta_n^2e^{-2q_0t_r}\right)-\Delta_n \left(1-e^{-2q_0t_r}\right)e^{-2q_n d_{1c}}\right], \\
&-\left[\Delta_n \left(1-e^{-2q_0t_l}\right)-\left( \Delta_n^2-e^{-2q_0t_l}\right)e^{-2q_n d_{1a}}\right]\left[\Delta_n \left(1-e^{-2q_0t_r}\right)-\left( \Delta_n^2-e^{-2q_0t_r}\right)e^{-2q_n d_{1c}}\right]e^{-2q_n d_{1b}},\\
q_n=&\sqrt{\xi^2+k_{\perp}^2+m_n^2},\hspace{1cm}\Delta_n=\frac{q_n-q_0}{q_n+q_0}.
\end{split}
\end{equation}It is easy to see that $\Xi^{\text{I}}(\xi,k_{\perp})$ goes to zero exponentially fast when $\xi\rightarrow \infty$ or $k_{\perp}\rightarrow \infty$. Therefore the integral in $\Lambda_{T,2}^{\text{I}}(s)$ is an analytic function of $s$ for all $s<1/2$. Because of the factor $\sin\pi s$, we find that $\Lambda_{T,2}(0)=0$. Therefore,
$$\zeta_T^{\text{cont, I}}(0)=\Lambda_{T,1}^{\text{I}}(0),\hspace{2cm}\zeta_T^{\text{cont, I} \;\prime}(0)=\Lambda_{T,1}^{\text{I}\;\prime}(0)+\Lambda_{T,2}^{\text{I}\;\prime}(0),$$and
the continuum mode contribution to the Casimir energy in the system I in FIG. \ref{f2} is
\begin{equation*}
\begin{split}
E_{\text{Cas}}^{\text{I, cont}}=-\frac{T}{2}\left(\zeta_T^{\text{cont, I} \;\prime}(0)+[\log \mu^2]\zeta_T^{\text{cont, I} }(0)\right).
\end{split}
\end{equation*}The corresponding Casimir energy for system II, $E_{\text{Cas}}^{\text{II, cont}}$, is obtained from $E_{\text{Cas}}^{\text{I, cont}}$ by replacing $d_{1a}, d_{1b}$ and $d_{1c}$ by $d_{2a}, d_{2b}$ and $d_{2c}$ respectively.  The contribution of the continuum modes to the renormalized Casimir energy of the parallel plate system is given by the limit
\begin{equation*}
\begin{split}
E_{\text{Cas}}^{\parallel, \text{cont}}=\lim_{\substack{d_{1a},d_{1c},d_{2a},d_{2b},d_{2c}\rightarrow\infty\\ d_{1b}=a\;\text{fixed}}}
\left(E_{\text{Cas}}^{\text{I, cont}}-E_{\text{Cas}}^{\text{II, cont}}\right).
\end{split}
\end{equation*} Since $d_{1a}+d_{1b}+d_{1c}=d_{2a}+d_{2b}+d_{2c}=L_1-t_l-t_r$, we find that $\Lambda_{T,1}^{\text{I}}(s)= \Lambda_{T,1}^{\text{II}}(s)$. Therefore, the contribution to the Casimir energy from the $\Lambda_{T,1}$ terms are the same for the two systems, and therefore cancel out after the subtraction.
Consequently,
\begin{equation}\label{eq8_4_9}
\begin{split}
E_{\text{Cas}}^{\parallel, \text{cont}}=&-\frac{T}{2}\lim_{\substack{d_{1a},d_{1c},d_{2a},d_{2b},d_{2c}\rightarrow\infty\\ d_{1b}=a\;\text{fixed}}}
\left(\Lambda_{T,2}^{\text{I}\;\prime}(0)-\Lambda_{T,2}^{\text{II}\;\prime}(0)\right)\\=& - \frac{AT}{2\pi}\sum_{n=0}^{\infty}\sum_{\ell=0}^{\infty}\!'\lim_{\substack{d_{1a},d_{1c},d_{2a},d_{2b},d_{2c}\rightarrow\infty\\ d_{1b}=a\;\text{fixed}}}\int_0^{\infty}\int_{2\pi\ell T}^{\infty}\frac{d}{d\xi}\ln \frac{\Xi^{\text{I}}(\xi,k_{\perp})}{\Xi^{\text{II}}(\xi,k_{\perp})}d\xi k_{\perp}dk_{\perp}\\
=&-\frac{AT}{2\pi}\sum_{n=0}^{\infty}\sum_{\ell=0}^{\infty}\!'\int_0^{\infty}\int_{2\pi\ell T}^{\infty}\frac{d}{d\xi}\ln\left\{1-\frac{\Delta_n^2 \left(1-e^{-2q_0t_l}\right)
\left(1-e^{-2q_0t_r}\right)}{\left(1-\Delta_n^2e^{-2q_0t_l}\right)\left(1-\Delta_n^2e^{-2q_0t_r}\right)}e^{-2q_n(\xi_{\ell},k_{\perp})a}\right\}d\xi k_{\perp}dk_{\perp}\\
=&\frac{AT}{2\pi}\sum_{n=0}^{\infty}\sum_{\ell=0}^{\infty}\!'\int_0^{\infty} \ln\left\{1-\frac{\Delta_n^2(\xi_{\ell},k_{\perp}) \left(1-e^{-2q_0(\xi_{\ell},k_{\perp})t_l}\right)
\left(1-e^{-2q_0(\xi_{\ell},k_{\perp}) t_r}\right)}{\left(1-\Delta_n(\xi_{\ell},k_{\perp}) ^2e^{-2q_0(\xi_{\ell},k_{\perp}) t_l}\right)\left(1-\Delta_n(\xi_{\ell},k_{\perp}) ^2e^{-2q_0(\xi_{\ell},k_{\perp}) t_r}\right)}e^{-2q_n(\xi_{\ell},k_{\perp})a}\right\}  k_{\perp}dk_{\perp},
\end{split}
\end{equation}where $\xi_{\ell}=2\pi\ell T$. Notice that
\begin{equation}\label{eq8_4_4}\Delta_n(\xi,k_{\perp})=\frac{q_n(\xi,k_{\perp})-q_0(\xi,k_{\perp})}{q_n(\xi,k_{\perp})+q_0(\xi,k_{\perp})}=\frac{\sqrt{\xi^2+k_{\perp}^2+m_n^2}-\sqrt{\xi^2+k_{\perp}^2}}
{\sqrt{\xi^2+k_{\perp}^2+m_n^2}+\sqrt{\xi^2+k_{\perp}^2}}. \end{equation} Therefore, $\Delta_0\equiv 0$ and we can start the summation over $n$ from $n=1$ instead of $n=0$. On the other hand,  \eqref{eq8_4_4} shows that  $0\leq \Delta_n(\xi,k_{\perp})<1$ for all real $\xi$ and $k_{\perp}$. From this, one can verify that
\begin{equation}\label{eq8_10_1} 0<\frac{\Delta_n^2(\xi,k_{\perp}) \left(1-e^{-2q_0(\xi,k_{\perp})t_l}\right)
\left(1-e^{-2q_0(\xi,k_{\perp}) t_r}\right)}{\left(1-\Delta_n(\xi,k_{\perp}) ^2e^{-2q_0(\xi,k_{\perp}) t_l}\right)\left(1-\Delta_n(\xi,k_{\perp}) ^2e^{-2q_0(\xi,k_{\perp}) t_r}\right)}\leq 1.\end{equation}Hence the continuum mode contribution to the Casimir energy of the parallel plate system $E_{\text{Cas}}^{\parallel, \text{cont}}$ is also always negative. The continuum mode contribution to the Casimir force is \begin{equation}\label{eq8_4_10}
\begin{split}
&F_{\text{Cas}}^{\parallel, \text{cont}}=-\frac{\pa E_{\text{Cas}}^{\parallel, \text{cont}}}{\pa a}
\\=&-\frac{AT}{\pi}\sum_{n=1}^{\infty}\sum_{\ell=0}^{\infty}\!'\int_0^{\infty}q_n(\xi_{\ell},k_{\perp})\left\{ \frac{\left(1-\Delta_n(\xi_{\ell},k_{\perp}) ^2e^{-2q_0(\xi_{\ell},k_{\perp}) t_l}\right)\left(1-\Delta_n(\xi_{\ell},k_{\perp}) ^2e^{-2q_0(\xi_{\ell},k_{\perp}) t_r}\right)}
{\Delta_n^2(\xi_{\ell},k_{\perp}) \left(1-e^{-2q_0(\xi_{\ell},k_{\perp})t_l}\right)
\left(1-e^{-2q_0(\xi_{\ell},k_{\perp}) t_r}\right)}e^{2q_n(\xi_{\ell},k_{\perp})a}-1\right\}^{-1}  k_{\perp}dk_{\perp},
\end{split}
\end{equation}which is always attractive. By taking the $T\rightarrow 0$ limit of \eqref{eq8_4_9} and \eqref{eq8_4_10} using \eqref{eq8_4_5}, we find that the continuum mode contribution to the zero temperature Casimir energy and zero temperature Casimir force are given respectively by
\begin{equation*}
\begin{split}
E_{\text{Cas}}^{\parallel, \text{cont}, T=0}=&\frac{A}{4\pi^2}\sum_{n=0}^{\infty} \int_0^{\infty}\int_0^{\infty} \ln\left\{1-\frac{\Delta_n^2(\xi ,k_{\perp}) \left(1-e^{-2q_0(\xi,k_{\perp})t_l}\right)
\left(1-e^{-2q_0(\xi,k_{\perp}) t_r}\right)}{\left(1-\Delta_n(\xi,k_{\perp}) ^2e^{-2q_0(\xi,k_{\perp}) t_l}\right)\left(1-\Delta_n(\xi,k_{\perp}) ^2e^{-2q_0(\xi,k_{\perp}) t_r}\right)}e^{-2q_n(\xi,k_{\perp})a}\right\} d\xi k_{\perp}dk_{\perp};
\end{split}
\end{equation*}and
\begin{equation*}
\begin{split}
&F_{\text{Cas}}^{\parallel, \text{cont}, T=0}\\=& -\frac{A}{2\pi^2}\sum_{n=1}^{\infty}\int_0^{\infty} \int_0^{\infty}q_n(\xi,k_{\perp})\left\{ \frac{\left(1-\Delta_n(\xi,k_{\perp}) ^2e^{-2q_0(\xi,k_{\perp}) t_l}\right)\left(1-\Delta_n(\xi,k_{\perp}) ^2e^{-2q_0(\xi,k_{\perp}) t_r}\right)}
{\Delta_n^2(\xi,k_{\perp}) \left(1-e^{-2q_0(\xi,k_{\perp})t_l}\right)
\left(1-e^{-2q_0(\xi,k_{\perp}) t_r}\right)}e^{2q_n(\xi,k_{\perp})a}-1\right\}^{-1}  d\xi k_{\perp}dk_{\perp}.
\end{split}
\end{equation*}

Collecting together the contribution from the discrete modes, we find that the Casimir force acting on a pair of parallel  plates under $4D$ perfectly conducting condition is given by
\begin{equation}\label{eq8_10_4}
\begin{split}
F_{\text{Cas},4D}^{\parallel}=&  F_{\text{Cas},0}^{\parallel}+2F_{\text{Cas},1}^{\parallel}+F_{\text{Cas}}^{\parallel, \text{cont}}.
\end{split}
\end{equation}Since each of the terms $F_{\text{Cas},0}^{\parallel}$, $F_{\text{Cas},1}^{\parallel}$ and $F_{\text{Cas}}^{\parallel, \text{cont}}$ is negative, we see that the Casimir force due to $4D$ perfectly conducting condition is always attractive and have larger magnitude than the Casimir force in $4D$ Minkowski spacetime. Compare to the Casimir force due to $5D$ induced perfectly conducting condition \eqref{eq7_30_4}, we find that their difference is
$$F_{\text{Cas},5D}^{\parallel}-F_{\text{Cas},4D}^{\parallel}=F_{\text{Cas},1}^{\parallel}-F_{\text{Cas}}^{\parallel, \text{cont}}.$$
The inequality \eqref{eq8_10_1} implies that
\begin{equation}\label{eq8_10_2}
\begin{split}
\left|F_{\text{Cas}}^{\parallel, \text{cont}}\right|=-F_{\text{Cas}}^{\parallel, \text{cont}}\leq \frac{AT}{\pi}\sum_{n=1}^{\infty}\sum_{\ell=0}^{\infty}\!'\int_0^{\infty}\frac{q_n(\xi_{\ell},k_{\perp}) }{ e^{2q_n(\xi_{\ell},k_{\perp})a}-1}  k_{\perp}dk_{\perp}.
\end{split}
\end{equation}It is elementary to verify that the right hand side of \eqref{eq8_10_2} is exactly equal to $-F_{\text{Cas},1}^{\parallel}=\left|F_{\text{Cas},1}^{\parallel}\right|$. Therefore, we find that the magnitude of the Casimir force due to $5D$ induced perfectly conducting condition \emph{is always larger than} the magnitude of the Casimir force due to $4D$  perfectly conducting  condition.

A fundamental difference between the Casimir effect due to $5D$ induced perfectly conducting condition and the Casimir force due to $4D$ perfectly conducting condition is that the former \emph{does not} depend on the thicknesses of the plates, but the later \emph{does}. Since $0\leq \Delta_n(\xi,k_{\perp})<1$ for all real $\xi$ and $k_{\perp}$, one can verify that
$$\frac{1-\Delta_n e^{-2q_0t}}{1-e^{-2q_0t}}$$ is a decreasing function of $t$. Hence, the magnitude of the continuum mode contribution to the Casimir force due to $4D$ perfectly conducting boundary condition is an increasing function of the plate thicknesses. In fact, it is easy to see that \emph{as the thicknesses of the plates goes to zero, the continuum mode contribution to the Casimir force goes to zero.} As the thicknesses of the plates goes to infinity, the continuum mode contribution to the Casimir force  tends to the limiting value
$$-\frac{AT}{\pi}\sum_{n=1}^{\infty}\sum_{\ell=0}^{\infty}\!'\int_0^{\infty}\frac{q_n(\xi_{\ell},k_{\perp}) }{ \Delta_n(\xi_{\ell},k_{\perp})^{-2}e^{2q_n(\xi_{\ell},k_{\perp})a}-1}  k_{\perp}dk_{\perp}.$$

\section{Numerical analysis}\label{s5}
\begin{figure}[h]\centering \epsfxsize=.49\linewidth
\epsffile{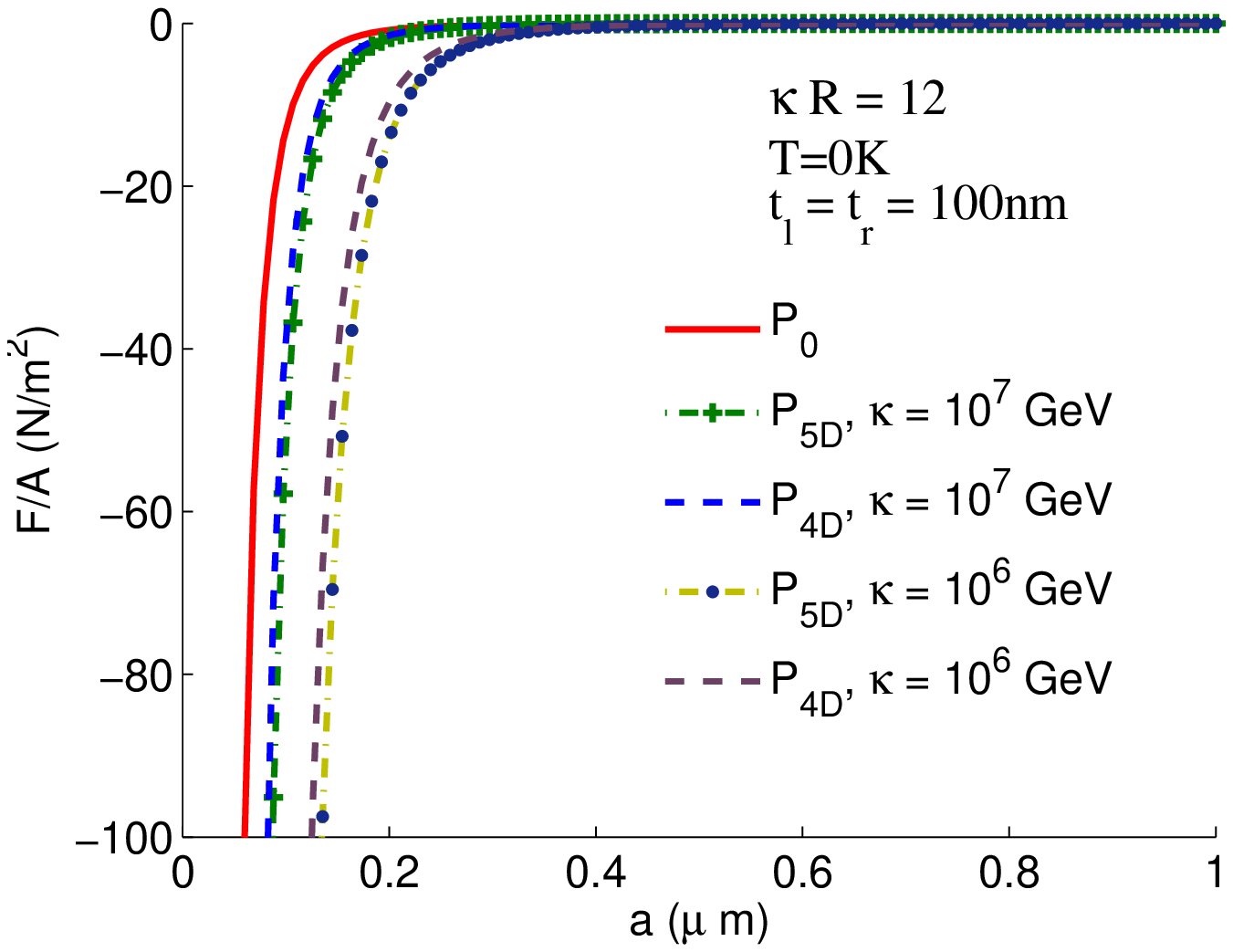}\centering \epsfxsize=.49\linewidth
\epsffile{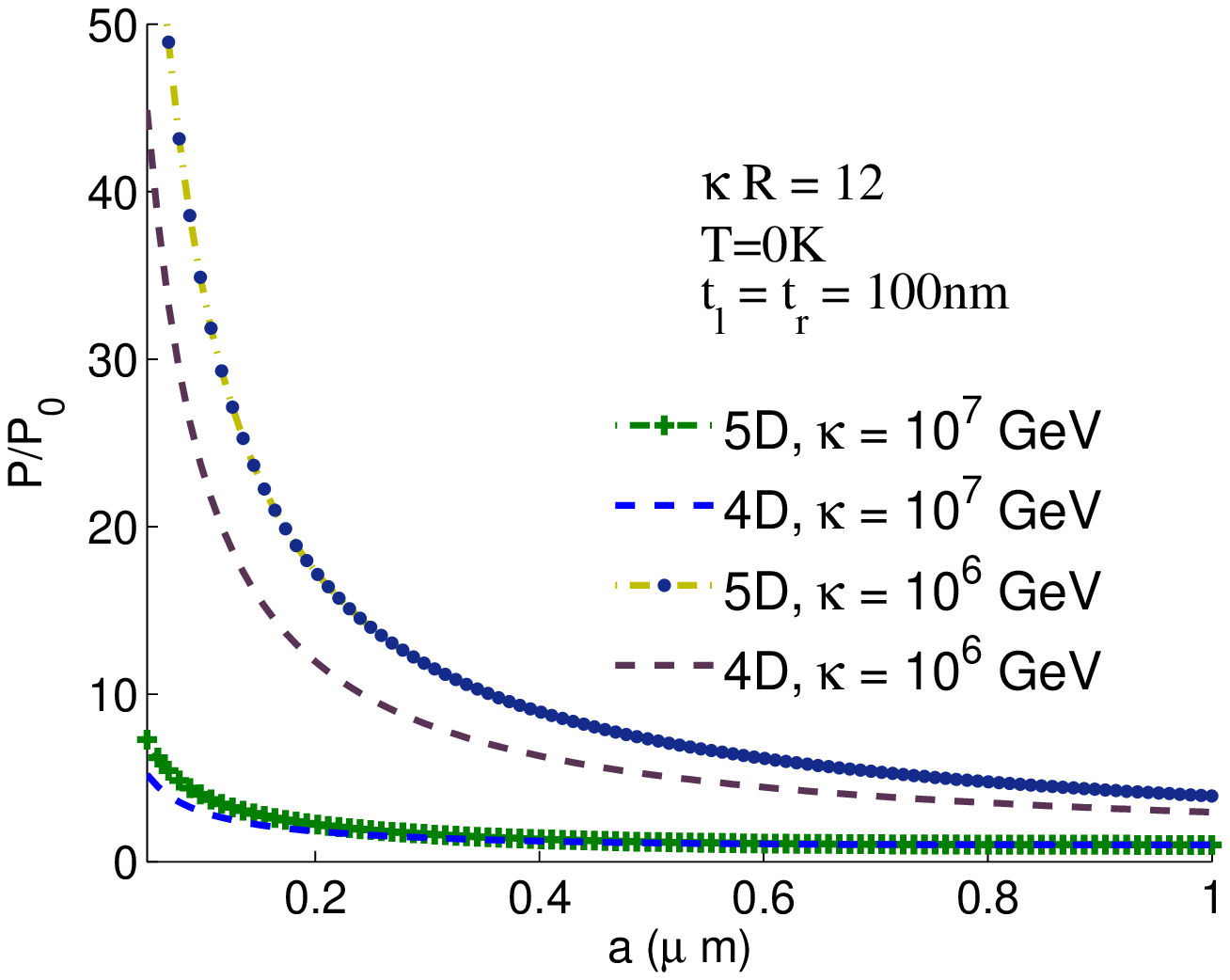}\caption{\label{f3} Comparisons between the Casimir pressure in the absence of  extra dimension ($P_0$), the Casimir pressure with $5D$ induced perfectly conducting condition ($P_{5D}$), and the Casimir pressure with $4D$ perfectly conducting condition ($P_{4D}$). Here $T = 0$K, $t_l=t_r=100$nm, $\kappa R=12$, $\kappa=10^7$GeV or $10^6$GeV.}\end{figure}

\begin{figure}[h]\centering \epsfxsize=.49\linewidth
\epsffile{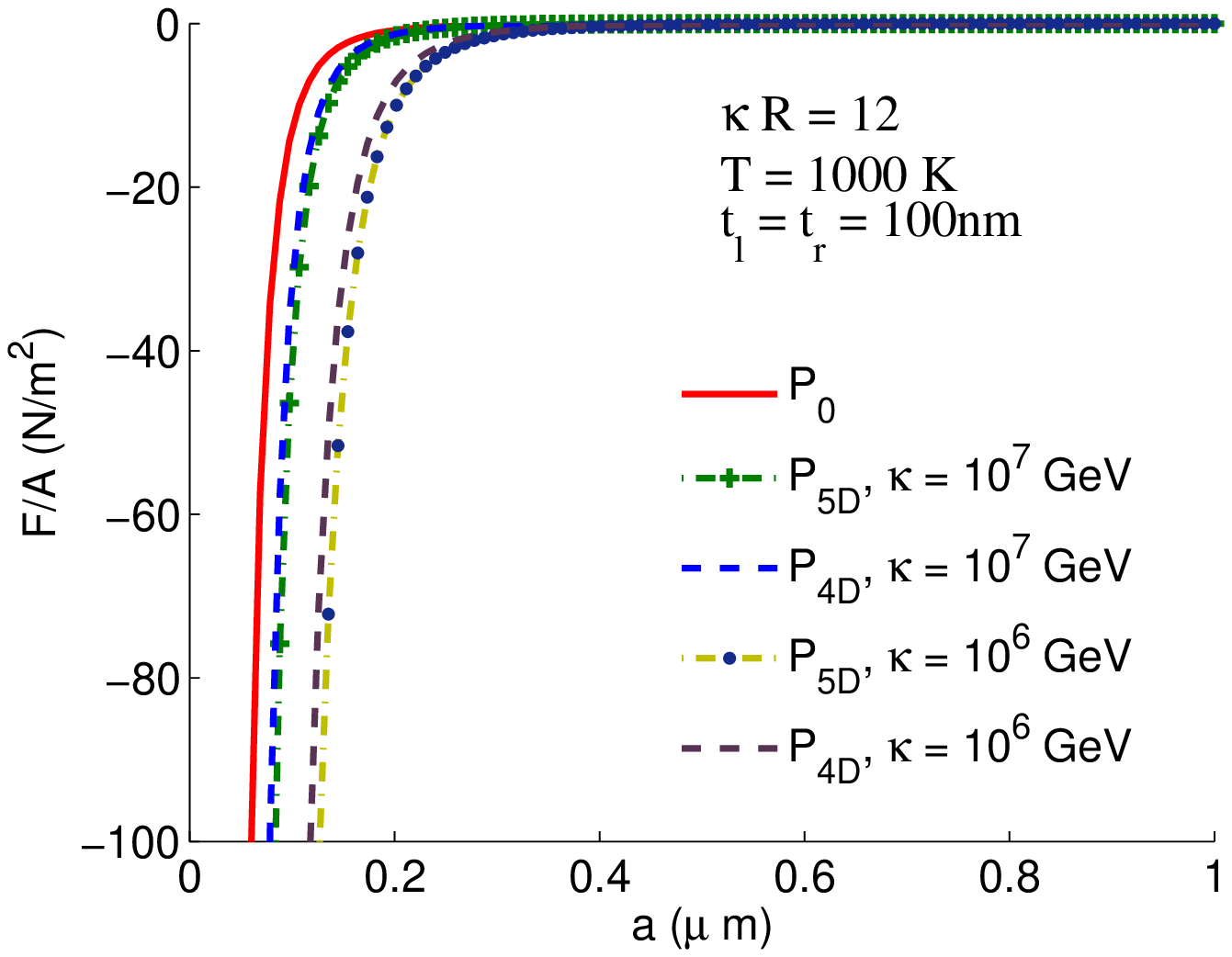}\centering \epsfxsize=.49\linewidth
\epsffile{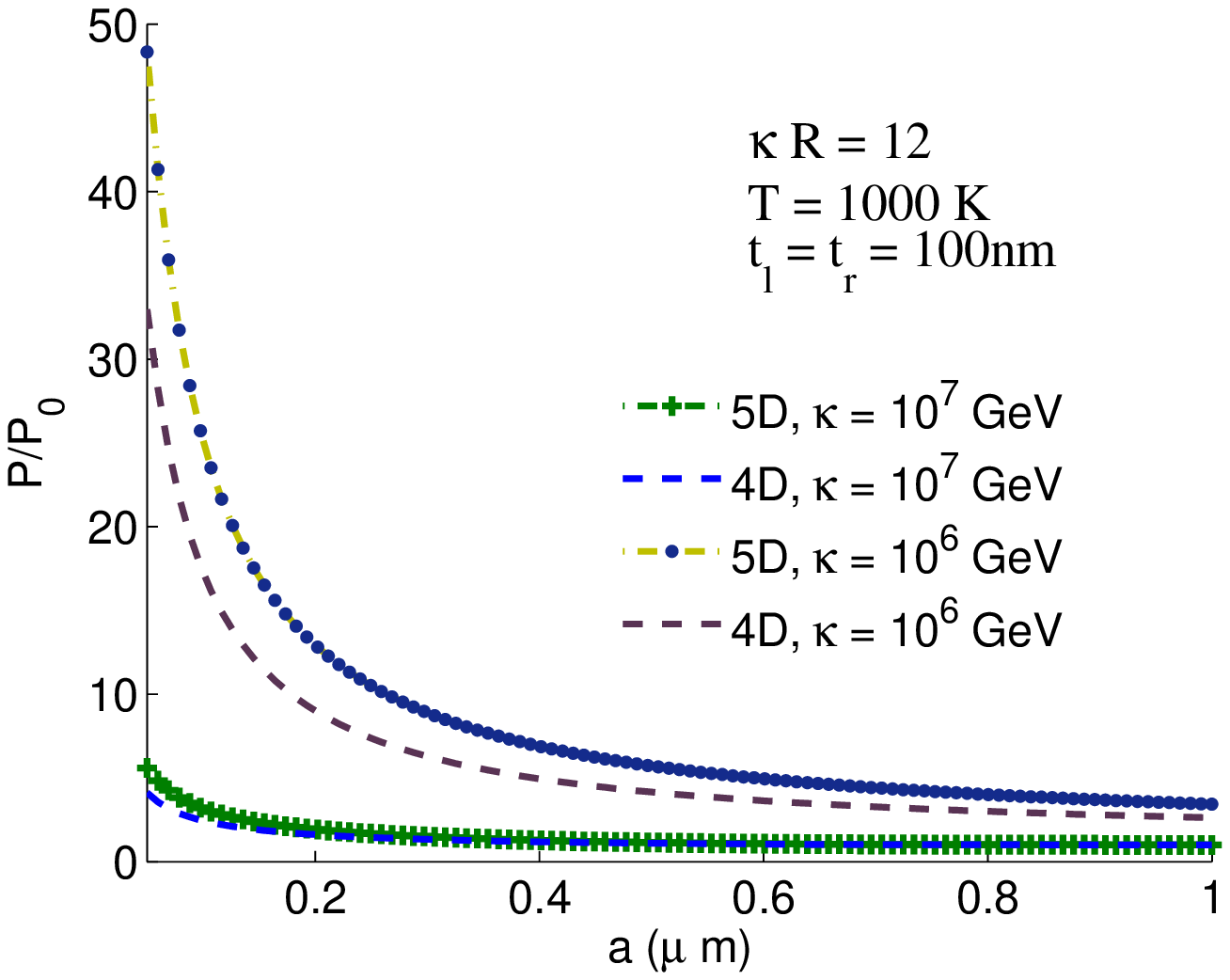}\caption{\label{f4} Same as FIG. \ref{f3} except $T=1000$K.}\end{figure}

From Section \ref{s3} and Section \ref{s4}, we find that the Casimir force acting on a pair of   parallel plates in RS model can be written as
\begin{align*}
F_{\text{Cas}, 5D}^{\parallel}= F_{\text{Cas},0}^{\parallel}+3 F_{\text{Cas},1}^{\parallel}\end{align*}for 5D induced perfectly conducting condition, and as
\begin{align*}
F_{\text{Cas}, 4D}^{\parallel}= F_{\text{Cas},0}^{\parallel}+2 F_{\text{Cas},1}^{\parallel}+  F_{\text{Cas}}^{\parallel, \text{cont}},
\end{align*}for $4D$ perfectly conducting boundary condition. Here $F_{\text{Cas},0}^{\parallel}$ is the Casimir force acting on a pair of parallel perfectly conducting plates in $4D$ Minkowski spacetime, $F_{\text{Cas},1}^{\parallel}$ is the correction to the Casimir force due to one discrete mode, and $F_{\text{Cas}}^{\parallel, \text{cont}}$ is the correction to the Casimir force due to a continuum mode.

In FIG. \ref{f3} and FIG. \ref{f4}, we show graphically the dependence of the Casimir forces on the plate separation $a$. We compare the Casimir force due to $5D$ induced perfectly conducting condition and the Casimir force due to $4D$ perfectly conducting condition to the Casimir force in $4D$ Minkowski spacetime. The Kaluza-Klein masses $m_n$ are computed numerically using bisection method. It is well-known that RS model was first proposed to solve the hierarchy problem between the Planck and electroweak scales and this requires $\kappa R\simeq 12$. For $\kappa R=12$ and the separation of the plates in the range 50nm to 1$\mu$m, we find that the Casimir force in RS model differ considerably ($> 1\%$) only if $\kappa < 10^{9}$ GeV. In FIG. \ref{f3}, we plot the graphs of the Casimir forces per unit area when $\kappa = 10^7$ GeV and $\kappa = 10^6$ GeV for $T=0$K. In FIG. \ref{f4}, we plot the graphs when $T=1000$K.
Numerical calculations show that the continuum mode correction $ F_{\text{Cas}}^{\parallel, \text{cont}}$ is very insignificant compared to the discrete mode correction $ F_{\text{Cas},1}^{\parallel}$ when $\kappa\geq 10^6$ GeV. The continuum mode correction is at least $30$ times smaller. Therefore, the correction to the Casimir force comes principally from the discrete mode correction. Under $5D$ induced perfectly conducting condition, there are three discrete mode corrections. Under $4D$ perfectly conducting condition, there are only two. Numerically, we find that at $\kappa = 10^8$ GeV, the ratio of the discrete mode correction $F_{\text{Cas},1}^{\parallel}$ to   $F_{\text{Cas},0}^{\parallel}$ is $\sim 6\%$ when $a=50$nm. Therefore, RS scenario gives  $\sim 18\%$ correction to the $4D$ Casimir force if one considers $5D$ induced perfectly conducting condition, and  $\sim 12\%$ correction if one considers $4D$ perfectly conducting condition. In fact, the corrections become larger when $\kappa$ gets smaller. For example, when $\kappa=10^6$ GeV, the correction due to one discrete mode becomes $\sim 2000\%$. Compare FIG. \ref{f4}   to FIG. \ref{f3}, it is interesting to note that the increase of temperature can reduce the percentage of correction.

\section{Perturbation of Casimir force by a noncommutativity parameter}\label{s6}
In this section, we briefly comment on the effect of spacetime noncommutativity on the sign of   Casimir force. For simplicity, we only consider the zero temperature case here. As discussed in \cite{47,48,18}, in the simplest case, spacetime noncommutativity can lead to a modification of the zero temperature Casimir energy by
\begin{equation*}
E_{\text{Cas}}^{T=0}=\frac{1}{2}\sum_{\text{modes}}\omega \quad \Longrightarrow \quad  E_{\text{Cas}}^{T=0}=\frac{1}{2}\sum_{\text{modes}}\omega e^{-\ell_{\text{nc}}^2\omega^2},
\end{equation*}where $\ell_{\text{nc}}$ is the fundamental noncommutative length scale. The zero temperature Casimir effect of scalar field in noncommutative RS model has been considered in \cite{18}, where the authors calculated the first order correction to the Casimir force. In \cite{49}, we used another method which allows us to compute to all orders of the noncommutative parameter $\ell_{\text{nc}}^2$.

For electromagnetic field with $5D$ induced perfectly conducting conditions, the effect of spacetime noncommutativity can be easily read from our previous result in \cite{49} by taking $p'=h_0=2$, $p=h=3$, which gives
\begin{equation} \label{eq8_4_1}
\begin{split}
&F_{\text{Cas}, 5D}^{\parallel, T=0} =-\frac{Ah_0}{4\pi^{\frac{5}{2}}}\sum_{j=0}^{\infty}\ell_{\text{nc}}^{2j}\frac{(j+1)}{a^{2j+4}}\Gamma\left(j+\frac{5}{2}\right)\zeta_R(2j+4)-\frac{Ah}
{4\pi^{\frac{5}{2}}}\sum_{j=0}^{\infty}\ell_{\text{nc}}^{2j}\frac{\Gamma\left(j+\frac{3}{2}\right)}{j!}\\&\times\left\{\frac{1}{a^{j+2}}\sum_{k=1}^{\infty}\sum_{n=1}^{\infty} (j+1) \left(\frac{m_n}{ k}\right)^{j+2}K_{j+2}(2 kam_n)\right.+\left.\frac{1}{a^{j+1}}\sum_{k=1}^{\infty}\sum_{n=1}^{\infty} \frac{m_n^{j+3}}{k^{j+1}}\left(K_{j+1}(2 kam_n)+K_{j+3}(2 kam_n)\right)\right\}.
\end{split}
\end{equation}As explained in \cite{49}, the functions $\Gamma(z), K_{\nu}(z)$ is always positive for $z>0$, and the Riemann zeta function $\zeta_R(z)$ is always positive for $z>1$. Therefore, the Casimir force is always negative (attractive) to any orders of the perturbation parameter $\ell_{\text{nc}}^2$.

Next, we consider the $4D$ perfectly conducting condition. As discussed in Section \ref{s4}, the energy eigenmodes can be divided into two discrete modes and one continuum mode. The contribution to the Casimir force from the discrete modes can be obtained from the Casimir force under $5D$ induced perfectly conducting condition \eqref{eq8_4_1}  by replacing $h=3$ with $h_0=2$. It is easily seen from \eqref{eq8_4_1} that this discrete mode contribution is also always negative (attractive) to any orders of the perturbation parameter $\ell_{\text{nc}}^2$. For the continuum mode contribution, we use the same approach as in \cite{49}. Using zeta regularization, we find that
\begin{equation*}\begin{split}
E_{\text{Cas}}^{T=0}=&\left.\frac{\mu^{2s}}{2}\sum_{\text{modes}}\omega^{1-2s} e^{-\ell_{\text{nc}}^2\omega^2}\right|_{s=0}\\
=&\left.\frac{1}{2}\sum_{j=0}^{\infty}\frac{(-1)^j}{j!}\ell_{\text{nc}}^{2j}\mu^{2s}\sum_{\text{modes}}\omega^{2j+1-2s}  \right|_{s=0}\\
=&\left.\frac{1}{2}\sum_{j=0}^{\infty}\frac{(-1)^j}{j!}\ell_{\text{nc}}^{2j}\mu^{2s}\zeta\left(s-j-\frac{1}{2}\right)\right|_{s=0}\\
=& \frac{1}{2}\sum_{j=0}^{\infty}\frac{(-1)^j}{j!}\ell_{\text{nc}}^{2j}\left(\text{FP}_{s=-j-\frac{1}{2}}\zeta(s)+[\log\mu^2]\text{Res}_{s=-j-\frac{1}{2}}\zeta(s)\right)\\
=&\sum_{j=0}^{\infty}\ell_{\text{nc}}^{2j}E_{\text{Cas},j}^{T=0},
\end{split}\end{equation*}where the $j^{\text{th}}$ order term is
$$E_{\text{Cas},j}^{T=0}=\frac{(-1)^j}{2j!}\left(\text{FP}_{s=-j-\frac{1}{2}}\zeta(s)+[\log\mu^2]\text{Res}_{s=-j-\frac{1}{2}}\zeta(s)\right).$$
Since the zeta function of a system $\zeta(s)$ \eqref{eq8_4_2} can be considered as taking only the $\ell=0$ term in the finite temperature zeta function of the system $\zeta_T(s)$ \eqref{eq8_4_2}, it is immediate to   obtain from Section \ref{s4} that in the system I (see FIG. \ref{f2}), the zeta function of the continuum modes $\zeta^{\text{cont, I}}(s)$ is given by
\begin{equation*}
\begin{split}
\zeta^{\text{cont, I}}(s)=&\Lambda_1^{\text{I}}(s)+\Lambda_2^{\text{I}}(s),
\end{split}
\end{equation*}where
\begin{equation*}
\begin{split}
\Lambda_1^{\text{I}}(s)=\frac{A}{ 2\pi^2}\sum_{n=0}^{\infty} \int_0^{\infty}\left\{(d_{1a}+d_{1b}+d_{1c})\int_{\sqrt{k_{\perp}^2+m_n^2}}^{\infty}
\frac{ x^{-2s}}{\sqrt{x^2-k_{\perp}^2-m_n^2}}dx+(t_l+t_r)\int_{ k_{\perp} }^{\infty}
\frac{ x^{-2s}}{\sqrt{x^2-k_{\perp}^2} }dx\right\}k_{\perp}dk_{\perp},\end{split}
\end{equation*}and\begin{equation*}
\begin{split}\Lambda_2^{\text{I}}(s)=&\frac{A\sin\pi s}{ 2\pi^2}\sum_{n=0}^{\infty} \int_0^{\infty}\int_{0}^{\infty} \xi^{-2s}\frac{d}{d\xi}\ln \Xi^{\text{I}}(\xi,k_{\perp})d\xi k_{\perp}dk_{\perp},\end{split}
\end{equation*}where $
\Xi^{\text{I}}(\xi,k_{\perp})$ is defined in \eqref{eq8_5_4}. The same reasoning as in Section \ref{s4} shows that $\Lambda_2^{\text{I}}(s)$ is analytic for $s<1/2$. Therefore, for any nonpositive integer $j$,
\begin{equation*}
\begin{split}
\text{Res}_{s=-j-\frac{1}{2}}\zeta^{\text{cont, I}}(s)=&\text{Res}_{s=-j-\frac{1}{2}}\Lambda_1^{\text{I}}(s),\\
\text{FP}_{s=-j-\frac{1}{2}}\zeta^{\text{cont, I}}(s)=&\text{FP}_{s=-j-\frac{1}{2}}\Lambda_1^{\text{I}}(s)+ \Lambda_2^{\text{I}}\left(-j-\frac{1}{2}\right).
\end{split}
\end{equation*}The $j^{\text{th}}$-order term of the continuum mode contribution to the Casimir energy of the parallel plate system is then given by
\begin{equation*}
\begin{split}
E_{\text{Cas}, j}^{\parallel, \text{cont}, T=0}=&\lim_{\substack{d_{1a},d_{1c},d_{2a},d_{2b},d_{2c}\rightarrow\infty\\ d_{1b}=a\;\text{fixed}}}
\left(E_{\text{Cas},j}^{\text{I, cont},T=0}-E_{\text{Cas},j}^{\text{II, cont},T=0}\right)\\=&\frac{(-1)^j}{2j!}
\lim_{\substack{d_{1a},d_{1c},d_{2a},d_{2b},d_{2c}\rightarrow\infty\\ d_{1b}=a\;\text{fixed}}}
\left(\left[\text{FP}_{s=-j-\frac{1}{2}}\zeta^{\text{cont, I}}(s)-\text{FP}_{s=-j-\frac{1}{2}}\zeta^{\text{cont, II}}(s)\right]\right.\\&\hspace{4cm}\left.+[\log\mu^2]\left[\text{Res}_{s=-j-\frac{1}{2}}\zeta^{\text{cont, I}}(s)-\text{Res}_{s=-j-\frac{1}{2}}\zeta^{\text{cont, II}}(s)\right]\right).
\end{split}
\end{equation*} As in Section \ref{s4},   $d_{1a}+d_{1b}+d_{1c}=d_{2a}+d_{2b}+d_{2c}=L_1-t_l-t_r$ implies that the contributions from the terms $\Lambda_{1}^{\text{I}}$ and  $\Lambda_{1}^{\text{II}}$ are the same for the two systems, and therefore cancel out after the subtraction.
Therefore, the $j^{\text{th}}$-order term of the continuum mode contribution to the Casimir energy of the parallel plate system is
\begin{equation*}
\begin{split}
E_{\text{Cas}, j}^{\parallel, \text{cont}, T=0}=&\frac{(-1)^j}{2j!}
\lim_{\substack{d_{1a},d_{1c},d_{2a},d_{2b},d_{2c}\rightarrow\infty\\ d_{1b}=a\;\text{fixed}}}\left(\Lambda_2^{\text{I}}\left(-j-\frac{1}{2}\right)-\Lambda_2^{\text{II}}\left(-j-\frac{1}{2}\right)\right)\\
=&-\frac{A}{4\pi^2 j!}\sum_{n=1}^{\infty} \int_0^{\infty}\int_{0}^{\infty}\xi^{2j+1}\frac{d}{d\xi}\ln\left\{1-\frac{\Delta_n^2 \left(1-e^{-2q_0t_l}\right)
\left(1-e^{-2q_0t_r}\right)}{\left(1-\Delta_n^2e^{-2q_0t_l}\right)\left(1-\Delta_n^2e^{-2q_0t_r}\right)}e^{-2q_n a}\right\}d\xi k_{\perp}dk_{\perp}\\
=& \frac{A(2j+1)}{4\pi^2 j!}\sum_{n=1}^{\infty} \int_0^{\infty}\int_{0}^{\infty}\xi^{2j } \ln\left\{1-\frac{\Delta_n^2 \left(1-e^{-2q_0t_l}\right)
\left(1-e^{-2q_0t_r}\right)}{\left(1-\Delta_n^2e^{-2q_0t_l}\right)\left(1-\Delta_n^2e^{-2q_0t_r}\right)}e^{-2q_n a}\right\}d\xi k_{\perp}dk_{\perp};
\end{split}
\end{equation*}and the $j^{\text{th}}$-order term of the continuum mode contribution to the Casimir force acting on the parallel plates is
\begin{equation*}
\begin{split}
F_{\text{Cas}, j}^{\parallel, \text{cont}, T=0}=&-\frac{\pa E_{\text{Cas}, j}^{\parallel, \text{cont}, T=0}}{\pa a}
\\=&-\frac{A(2j+1)}{2\pi^2 j!}\sum_{n=1}^{\infty} \int_0^{\infty}\int_{0}^{\infty}\xi^{2j } q_n \left\{ \frac{\left(1-\Delta_n^2e^{-2q_0t_l}\right)\left(1-\Delta_n^2e^{-2q_0t_r}\right)}{\Delta_n^2 \left(1-e^{-2q_0t_l}\right)
\left(1-e^{-2q_0t_r}\right)}e^{ 2 q_na}-1\right\}^{-1}d\xi k_{\perp}dk_{\perp},
\end{split}
\end{equation*}which is always (negative) attractive.

In conclusion, we find that to any order of the noncommutative parameter $\ell_{\text{nc}}^2$, the Casimir force on parallel perfectly conducting  plates is always attractive, whether we consider $5D$ induced or $4D$ perfectly conducting boundary conditions.
\section{Conclusion}

In this article, we compute the finite temperature Casimir force acting on a pair of parallel perfectly conducting plates in Randall-Sundrum model. Contrary to the previous related works, we do not use scalar field analogy here. There are two ways to interpret perfectly conducting condition in this model. One is induced from the $5D$ perfectly conducting condition. The other one that we call $4D$ perfectly conducting condition requires dimensional reduction to decompose the electromagnetic field in the $5D$ model into a Kaluza-Klein zero mode and a tower of Kaluza-Klein excitation modes, treated as $4D$ Maxwell field for massless photons and $4D$ Proca fields for massive photons respectively. We have shown that the $5D$ induced perfectly conducting condition and the $4D$ perfectly conducting condition give rise to different Casimir effect. Under $5D$ induced perfectly conducting condition, the Casimir force in RS model is the sum of the $4D$ Casimir force and three discrete mode corrections. Under $4D$ perfectly conducting condition, the Casimir force in RS model is the sum of the $4D$ Casimir force with two discrete mode corrections and one continuum mode correction.
Although our zero temperature  Casimir force for $5D$ induced perfectly conducting condition is similar to the result of \cite{14}, we would like to emphasize that the Kaluza-Klein masses for electromagnetic field are different from the Kaluza-Klein masses for scalar field used in \cite{14}.

We have shown that the magnitude of a continuum mode correction to the Casimir force is always less than the magnitude of a discrete mode correction to the Casimir force. Therefore, the magnitude of the Casimir force due to $4D$ perfectly conducting condition is always less  than the magnitude of the Casimir force due to $5D$ induced perfectly conducting condition. Numerically, the continuum mode correction to the Casimir force is much smaller than the discrete mode correction. Therefore, there is a significant difference between the Casimir forces under $5D$ induced condition and under $4D$ condition. It is interesting to note that the continuum mode contribution to the Casimir force under $4D$ perfectly conducting condition depends on the thicknesses of the plates. It goes to zero when the thicknesses of the plates go to zero, and it increases to a limiting value when the thicknesses increase.

Under either $5D$ induced or $4D$ perfectly conducting conditions, we find that the corrections to the $4D$ Casimir force always increase the magnitude of the attractive Casimir force. A brief  section is devoted to the study of  perturbation of  the Casimir force by a noncommutative parameter. It is established that the  perturbation due to noncommutativity  does not change the attractive nature of the Casimir force.

\end{document}